\documentclass[aps,prb,twocolumn,eqsecnum]{revtex4-2}

\usepackage{graphicx}% Include figure files
\usepackage{bm}% bold math
\usepackage{amsmath}
\usepackage{amssymb}

\newcommand{\ocite}{\onlinecite}

\newcommand{\iy}{\infty}

\newcommand{\x}{\text}
\newcommand{\pd}{\partial}

\newcommand{\dg}{\dagger}
\newcommand{\lan}{\langle}
\newcommand{\ran}{\rangle}
\newcommand{\lt}{\left}
\newcommand{\rt}{\right}

\newcommand{\f}{\frac}
\newcommand{\tf}{\tfrac}

\newcommand{\sq}{\sqrt}
\newcommand{\lbl}{\label}

\newcommand{\nn}{\nonumber}
\newcommand{\cd}{\cdot}

\newcommand{\p}{\perp}

\newcommand{\nm}{\hat{0}}
\newcommand{\um}{\hat{1}}

\newcommand{\tm}{\times}
\newcommand{\ot}{\otimes}

\newcommand{\eq}[1]{Eq.~(\ref{eq:#1})}
\newcommand{\eqs}[2]{Eqs.~(\ref{eq:#1}) and (\ref{eq:#2})}
\newcommand{\eqss}[3]{Eqs.~(\ref{eq:#1}), (\ref{eq:#2}), and (\ref{eq:#3})}

\newcommand{\eqn}[1]{(\ref{eq:#1})}
\newcommand{\eqsn}[2]{(\ref{eq:#1}) and (\ref{eq:#2})}

\newcommand{\secr}[1]{Sec.~\ref{sec:#1}}
\newcommand{\secsr}[2]{Secs.~\ref{sec:#1} and \ref{sec:#2}}

\newcommand{\figr}[1]{Fig.~\ref{fig:#1}}
\newcommand{\figsr}[2]{Figs.~\ref{fig:#1} and \ref{fig:#2}}
\newcommand{\appr}[1]{Appendix~\ref{app:#1}}

\newcommand{\spc}{\mbox{ }}

\newcommand{\beq}{\begin{equation}}
\newcommand{\eeq}{\end{equation}}
\newcommand{\beqar}{\begin{eqnarray}}
\newcommand{\eeqar}{\end{eqnarray}}
\newcommand{\beqarn}{\begin{eqnarray*}}
\newcommand{\eeqarn}{\end{eqnarray*}}
\newcommand{\ba}{\begin{array}}
\newcommand{\ea}{\end{array}}
\newcommand{\bwt}{\begin{widetext}}
\newcommand{\ewt}{\end{widetext}}

\newcommand{\ua}{\uparrow}
\newcommand{\da}{\downarrow}

\newcommand{\lrarr}{\leftrightarrow}

\newcommand{\rarr}{\rightarrow}

\newcommand{\dx}{{\text d}}
\newcommand{\ex}{{\text e}}

\newcommand{\ix}{{\text i}}

\newcommand{\Tx}{{\text T}}
\newcommand{\Ux}{{\text U}}

\newcommand{\ch}{\hat{c}}

\newcommand{\kh}{\hat{k}}

\newcommand{\ph}{\hat{p}}

\newcommand{\uh}{\hat{u}}
\newcommand{\vh}{\hat{v}}

\newcommand{\Ch}{\hat{C}}

\newcommand{\Hh}{\hat{H}}

\newcommand{\Th}{\hat{T}}
\newcommand{\Uh}{\hat{U}}

\newcommand{\Wh}{\hat{W}}
\newcommand{\Xh}{\hat{X}}

\newcommand{\epsh}{\hat{\varepsilon}}

\newcommand{\psih}{\hat{\psi}}

\newcommand{\chih}{\hat{\chi}}

\newcommand{\Deh}{\hat{\Delta}}

\newcommand{\Psih}{\hat{\Psi}}

\newcommand{\er}{{\bar{\epsilon}}}

\renewcommand{\vr}{{\bar{v}}}
\newcommand{\psirh}{\hat{\bar{\psi}}}

\newcommand{\Cc}{\mathcal{C}}

\newcommand{\Ec}{\mathcal{E}}

\newcommand{\Hc}{\mathcal{H}}

\newcommand{\Kc}{\mathcal{K}}

\newcommand{\Tc}{\mathcal{T}}

\newcommand{\Xc}{\mathcal{X}}

\newcommand{\Hch}{\hat{\Hc}}

\newcommand{\Xch}{\hat{\Xc}}

\newcommand{\at}{\tilde{a}}

\newcommand{\hb}{{\bf h}}

\newcommand{\kb}{{\bf k}}

\newcommand{\rb}{{\bf r}}

\newcommand{\Bb}{{\bf B}}
\newcommand{\Eb}{{\bf E}}

\newcommand{\al}{\alpha}
\newcommand{\be}{\beta}
\newcommand{\ga}{\gamma}

\newcommand{\de}{\delta}
\newcommand{\De}{\Delta}

\newcommand{\sig}{\sigma}
\newcommand{\Sig}{\Sigma}

\newcommand{\ka}{\varkappa}

\newcommand{\eps}{\varepsilon}
\newcommand{\e}{\epsilon}

\begin{document}
\title{Ever-present Majorana bound state in a generic one-dimensional superconductor with odd number of Fermi surfaces}

\author{Maxim Kharitonov$^{1,2}$, Ewelina M. Hankiewicz$^{1,3}$,  Bj\"orn Trauzettel$^{1,3}$, F. Sebastian Bergeret$^{2,4}$}
\address{$^1$Institute for Theoretical Physics and Astrophysics,
University of W\"urzburg, 97074 W\"urzburg, Germany\\
$^2$Donostia International Physics Center (DIPC),
Manuel de Lardizabal 4, E-20018 San Sebastian, Spain\\
$^3$W\"urzburg-Dresden Cluster of Excellence ct.qmat, Germany\\
$^4$Centro de F\'isica de Materiales (CFM-MPC), Centro Mixto CSIC-UPV/EHU,\\
Manuel de Lardizabal 5, E-20018 San Sebasti\'an, Spain
}

\begin{abstract}

A quasi-1D superconductor with odd number of Fermi surfaces
is expected to exhibit a nondegenerate Majorana bound state at the Fermi level
at its boundary with an insulator
(where the latter could be an actual insulator material or vacuum, for a terminated sample).
Previous explicit theoretical demonstrations of this property
were done for specific microscopic models of the bulk Hamiltonian and, most importantly, of the boundary.
In this work, we theoretically demonstrate that this property holds for the whole class of systems,
using the symmetry-based formalism of low-energy continuum models
and general boundary conditions.
We derive the general form of the Bogoliubov-de Gennes low-energy Hamiltonian
that is subject only to charge-conjugation symmetry $\Cc_+$ of the type $\Cc_+^2=+1$
and a few minimal assumptions.
Crucially, we also derive the most general form of the boundary conditions describing the boundary with an insulator,
subject only to the fundamental principle of the probability-current conservation and $\Cc_+$ symmetry.
Such {\em normal-reflection} boundary conditions do not contain scattering between electrons and holes.
We find that for odd number of Fermi surfaces a Majorana bound state always exists
as long as the bulk is in the gapped superconducting state,
irrespective of the parameters of the bulk Hamiltonian and boundary conditions.
Importantly, our general model includes a possible {\em Fermi-point mismatch},
when the two Fermi points are not at exactly opposite momenta, which disfavors superconductivity.
We find that the Fermi-point mismatch does {\em not} have a direct destructive effect on the Majorana bound state,
in the sense that once the bulk gap is opened the bound state is always present.

\end{abstract}
\maketitle

\section{Introduction and main results \lbl{sec:intro}}

\begin{figure*}
\includegraphics[width=.40\textwidth]{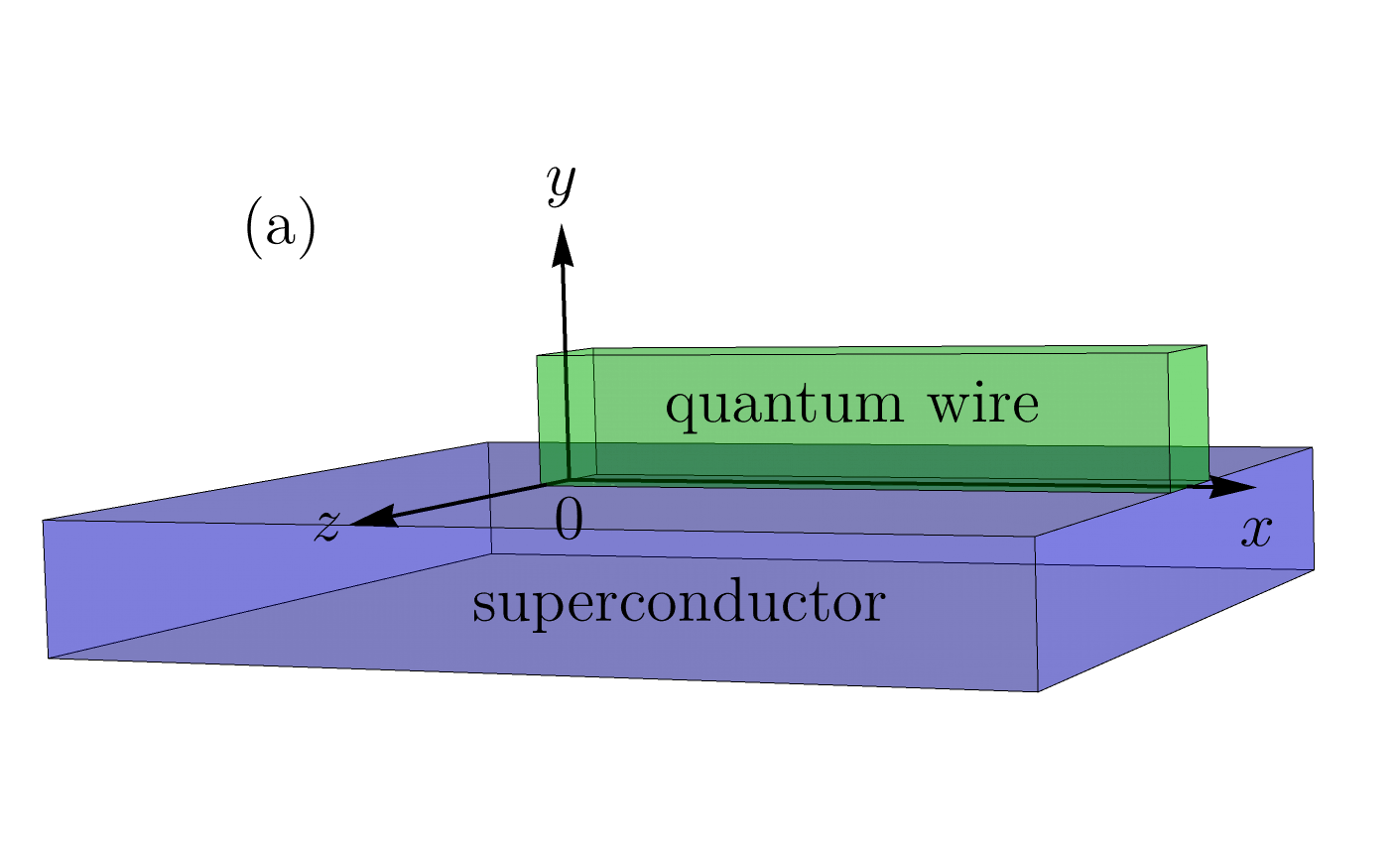}
\includegraphics[width=.35\textwidth]{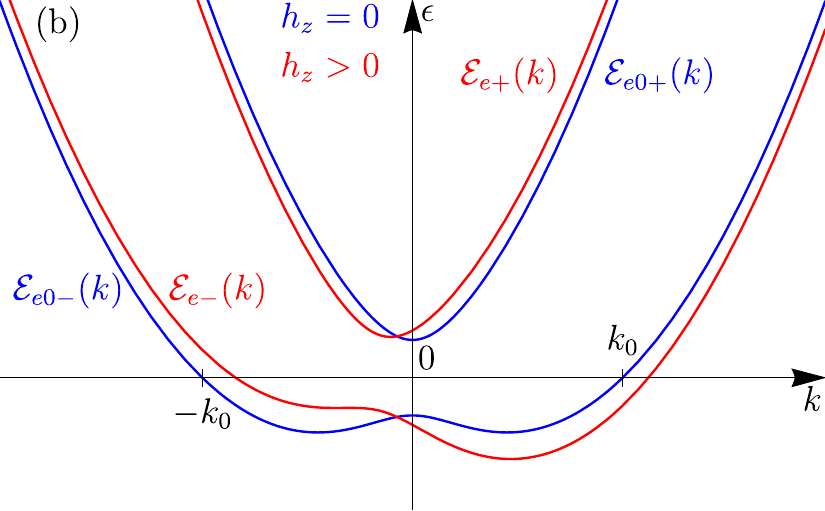}
\caption{
(a) The system of a quantum wire
coupled to a superconductor by the proximity effect~\cite{LdS,Oreg}, considered in \secr{qw}.
The Zeeman field is required to create the regime of 1FS.
Spin-orbit interactions are required to induce superconductivity
in the wire from the superconductor with spin-singlet pairing.
For the Zeeman field in the vertical plane containing the wire axis,
the electron system has~\cite{Sprb95,Sann,Samokhin} an effective time-reversal (TR) symmetry
with the operation $\Tc_{e+}=\Sig_z\Tc_{e-}$, $\Tc_{e+}^2=1$,
that is the product of the actual TR operation $\Tc_{e-}$, $\Tc_{e-}^2=-1$,
and reflection $\Sig_z$ along the horizontal direction $z$ perpendicular to the wire~\cite{Te}.
The component $h_z$ of the Zeeman field along the $z$ direction
breaks this $\Tc_{e+}$ symmetry and creates a {\em Fermi-point mismatch}
in the bulk normal-state electron spectrum $\Ec_\pm(k)$ [\eq{Ee}] (red), shown in (b).
The spectrum $\Ec_{e0\pm}(k)=\Ec_{e0\pm}(-k)$ [\eq{Ee0}] (blue) at $h_z=0$ is $k\lrarr-k$ symmetric due to the said $\Tc_{e+}$ symmetry.
}
\lbl{fig:qw}
\end{figure*}

\begin{figure*}
\includegraphics[width=.40\textwidth]{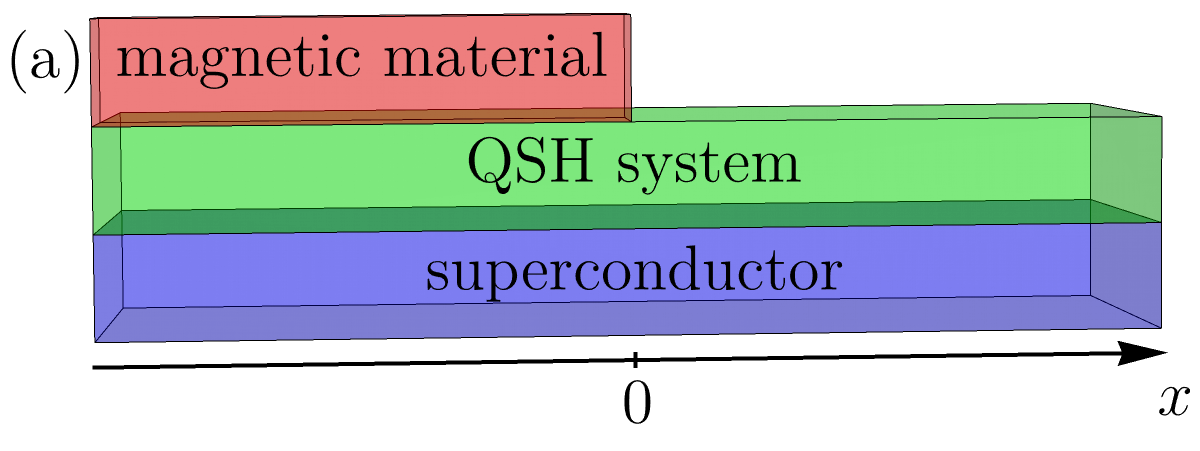}
\includegraphics[width=.27\textwidth]{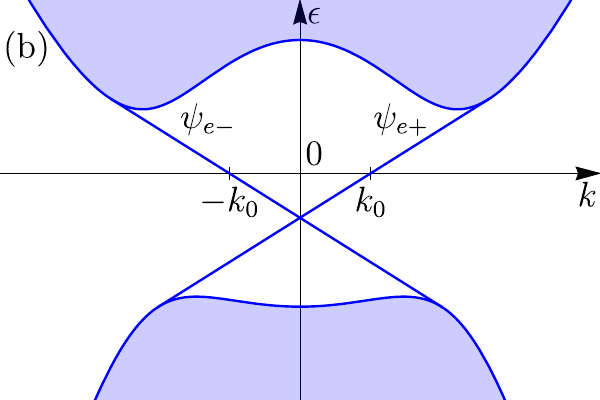}
\caption{
(a) The edge of a quantum spin Hall system coupled to a superconductor by the proximity effect~\cite{Fu}, considered in \secr{qsh}.
The magnetic material placed in the region $x<0$
breaks the time-reversal symmetry $\Tc_{e-}$, $\Tc_{e-}^2=-1$, of the electron system
and causes backscattering of the counterpropagating edge states.
(b) The schematic normal-state electron spectrum of the system, with counterpropagating edge states
and bulk continuum (shaded regions).
}
\lbl{fig:qsh}
\end{figure*}

Majorana bound states~\cite{Kitaev,Fu,LdS,Oreg,Potter,AkhmerovMprb,Wimmer,AkhmerovMprl,Lutchyn2011,
Fulga,Stanescu,Potter2011,Zhou,Law,LutchynFisher,
Yada,Sato,
Kells,Gibertini,
Rex,
Crepin,
Ikegaya,
Alicea,Leijnse,Tanaka,Beenakker,
Mourik,Churchill,Jack}
in quasi-one-dimensional superconductors
have attracted a lot of interest recently,
largely due to the prospects of implementing them as the basis for quantum computing.
Theoretical predictions for specific systems,
such as a quantum wire~\cite{LdS,Oreg} (\figr{qw}) or an edge of a quantum spin Hall system~\cite{Fu}
(\figr{qsh}) in proximity to a superconductor,
have been made and are being currently experimentally explored~\cite{Mourik,Churchill,Jack}.

A quasi-1D superconductor with odd number of Fermi surfaces (FSs)
(by a Fermi surface in 1D we mean a pair of Fermi points
with right- and left-moving electrons, in the geometry of \figr{system})
is expected on topological grounds
to exhibit a nondegenerate Majorana bound state at the Fermi level
at its boundary with an insulator
(where the latter could be an actual insulator material or vacuum, for a terminated sample).
Previous explicit theoretical demonstrations~\cite{Kitaev,Fu,LdS,Oreg,Potter,AkhmerovMprb,Wimmer,AkhmerovMprl,Lutchyn2011,
Fulga,Stanescu,Potter2011,Zhou,Law,LutchynFisher,
Kells,Gibertini,
Rex}
of this property
were done for specific microscopic models of the bulk Hamiltonian and, most importantly, of the boundary.

In this work, we theoretically demonstrate that this property holds for the whole class of systems,
using the symmetry-based formalism of low-energy continuum models
and general boundary conditions (BCs).
Remarkably, this formalism allows one to study the bound states of topological systems
in an explicit and general fashion,
while completely bypassing topological arguments, such as appeal to bulk-boundary correspondence.

In most of the paper, we perform such analysis for the case of one Fermi surface (1FS)
and then generalize this result to the case of arbitrary odd number of Fermi surfaces.

For the case of 1FS, under the approximation of the linearized
normal-state spectrum close to the Fermi points (\figr{system}),
we derive the most general form of the single-particle Bogoliubov--de Gennes~\cite{BdG} (BdG)
low-energy Hamiltonian that is subject only to charge-conjugation (CC) symmetry $\Cc_+$ of the type $\Cc_+^2=+1$.
Such system belongs to the symmetry class D~\cite{Chiu,Ryu};
physically, this describes spinful superconductors with no other assumed symmetries,
in particular, with broken time-reversal (TR) symmetry $\Tc_-$, $\Tc_-^2=-1$.

Crucially, for 1FS, we also derive the most general form of the BCs for this low-energy Hamiltonian,
subject only to the fundamental principle of the probability-current
conservation~\cite{Berry,Bonneau,Tokatly,McCann,AkhmerovPRL,AkhmerovPRB,Ostaay,Ahari,KharitonovLSM,KharitonovQAH,Seradjeh}
and to CC symmetry $\Cc_+$.
We find that there are two families of such BCs, which we term {\em normal-reflection} and {\em Andreev-reflection} BCs.
In the normal-reflection BCs, there is no scattering between electrons and holes,
electrons are reflected as electrons and holes as holes.
In the Andreev-reflection BCs,
there is complete reflection of electrons as holes and vice versa.

In this work, the physical systems of interest are superconductors interfaced with an insulator.
Such boundaries can only be described by normal-reflection BCs.
One the other hand, Andreev-reflection BCs cannot be defined in the normal state
and thus cannot represent the boundary with an insulator.
For this reason, we explore the bound states only for the normal-reflection BCs.
For the normal-reflection BCs,
we find that a single Majorana bound state
always exists at the boundary of a half-infinite system (\figr{gapped}),
as long as the bulk is in the gapped superconducting state,
irrespective of the parameters of the bulk Hamiltonian and BCs,
thereby proving our claim formulated above.

Importantly, our general model includes the possible {\em Fermi-point mismatch} (\figr{system}),
i.e., the situation when the two Fermi points are not at exactly opposite momenta~\cite{Nesterov,Rex},
which disfavors superconductivity and introduces a threshold (\figsr{gapped}{gapless})
for creating a gapped superconducting state by a coordinate-independent pairing field.
We also include the one-harmonic periodic coordinate dependence of the pairing field, which could help mitigate this effect.
We find that the Fermi-point mismatch does {\em not} have a direct destructive effect on the Majorana bound state,
in the sense that once the bulk gap is opened the bound state is always present.

We stress that our approach proves the existence of Majorana bound states {\em generally},
since this is demonstrated within the low-energy model of the most general form,
constrained only by the CC symmetry and the probability-current conservation principle.
The only requirement for the applicability of the low-energy model for 1FS
is the smallness of the superconducting pairing field and possible Fermi-point mismatch
compared to the nonlinearity scale of the underlying normal-state band
(these constraints can also be relaxed by the continuity argument).
In this low-energy limit, any microscopic BdG model will reduce
to an instance of the derived low-energy model with specific parameters
of its Hamiltonian and BCs. We illustrate such systematic procedure
of deriving the low-energy model from the microscopic model
with two examples: a generalized quantum-wire model (\figr{qw})
and a model of the edge of the quantum spin Hall system interfaced with a magnetic material (\figr{qsh}).

This claim thus holds for any microscopic BdG model,
regardless of the structure of its normal-state bulk Hamiltonian,
spin structure of the superconducting pairing field,
and boundary with an insulator.
Our low-energy symmetry-based approach thus proves that the persistence of Majorana bound states is, in fact,
the property of the whole class of systems.
This is the main difference from the previous theoretical demonstrations
of the existence of the Majorana bound states,
which were done for specific microscopic models.

The above approach for 1FS allows for a straightforward generalization to an arbitrary number of Fermi surfaces.
The only additional assumption we make
is to neglect superconducting pairing between different Fermi surfaces,
which is justified in the low-energy limit when the Fermi surfaces are well separated.
For more than 1FS, we derive from the start only the most general form of the normal-reflection BCs,
in which scattering between electrons and holes is absent,
since only such BCs describe the boundary with an insulator.
For odd number of Fermi surfaces,
we find that a Majorana bound state exists, irrespective of the parameters of the bulk Hamiltonian and BCs.

Our findings have crucial implications for the stability and persistence of Majorana bound states
in real systems and can be used as a general guide for engineering systems that host Majorana bound states:
as long as the basic general requirements of creating a system with odd number of Fermi surfaces
and inducing a gapped superconducting state in it are achieved, Majorana bound states are guaranteed to exist.

The paper is organized as follows.
All sections, except for \secr{NFS}, are devoted to the case of 1FS.
In \secr{H}, we derive the low-energy Hamiltonian of the most general form.
In \secr{bc}, we derive the BCs of the most general form.
In \secr{Deq}, we analyze the Fermi-point mismatch and introduce
the coordinate dependence of the pairing field that may help mitigate it.
In \secr{bs}, we derive our central result, the existence of the Majorana bound state.
In \secr{NFS}, we generalize this analysis to the case of arbitrary number of Fermi surfaces.
As examples of the microscopic realization of the general low-energy model for 1FS,
in \secr{qw}, we present the model of a quantum wire
and, in \secr{qsh},
the model of the edge of a quantum spin Hall system in proximity of the magnetic material.
In the concluding \secr{conclude},
we discuss the relation of our low-energy symmetry-based approach to studying bound states
to the topological aspect of the system and provide an outlook.
In \appr{T}, we analyze the effect of TR symmetries $\Tc_\pm$ with $\Tc_\pm^2=\pm1$.

Before we proceed, we mention that
during the preparation of the manuscript, Ref.~\ocite{Samokhin}
came out, where a similar main conclusion about the existence of Majorana bound states
was reached using a similar low-energy model.
We discuss the main differences between our work and Ref.~\ocite{Samokhin}
in \secr{Samokhin} after having presented our results.

\section{General low-energy Hamiltonian for one Fermi surface\lbl{sec:H}}

\begin{figure}
\includegraphics[width=.35\textwidth]{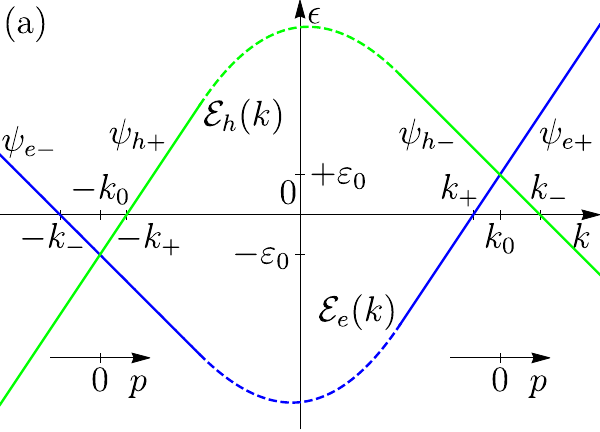}\\
\includegraphics[width=.45\textwidth]{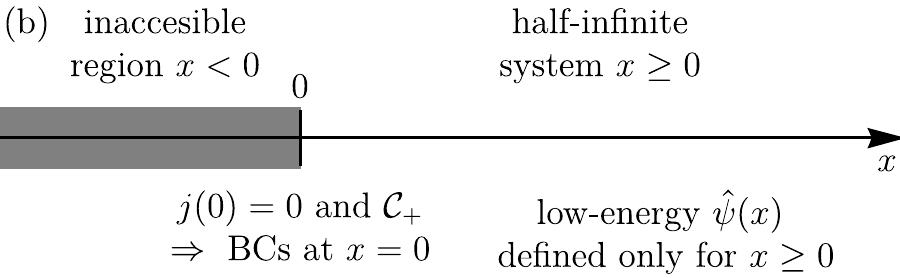}
\caption{
The origin of the generic low-energy model
of a 1D superconductor in the regime of one Fermi surface (1FS).
(a)
The schematic of the electron $\Ec_e(k)$ (blue) and hole $\Ec_h(k)=-\Ec_e(-k)$ (green) bulk bands
in the normal state, in the absence of superconducting pairing.
The solid-line parts show regions of the linearized spectrum
close to the Fermi level, where the low-energy model [\eqs{He}{H}] applies,
with the electron $\psi_{e\pm}(x)$ and hole $\psi_{h\pm}(x)$ components
of the BdG wave function $\psih(x)$ [\eq{psi}].
Generally, $\Ec_e(k)\neq \Ec_e(-k)$ and a {\em Fermi-point mismatch}
is present, when the Fermi points $\pm k_\pm$
are not at opposite momenta, $k_+\neq k_-$.
In this case, the electron $\Ec_e(k)$ and hole $\Ec_h(k)$ bands
cross at opposite momenta $\pm k_0$ at finite mismatch energies $\pm\eps_0$ [\eq{e0}]
relative to the Fermi level.
(b) Half-infinite low-energy system $x\geq 0$ with a boundary $x=0$ used to calculate bound states.
The low-energy BdG wave function  $\psih(x)$
is not defined in the ``inaccessible'' region $x<0$.
Instead, the general boundary conditions [\eqs{bcNR}{bcAR}] at $x=0$
that satisfy only the fundamental probability-current conservation principle [\eq{j=0}]
and charge-conjugation (CC) symmetry $\Cc_+$ are imposed (\secr{bc}, \figr{bc}).
}
\lbl{fig:system}
\end{figure}

\begin{figure}
\includegraphics[width=.45\textwidth]{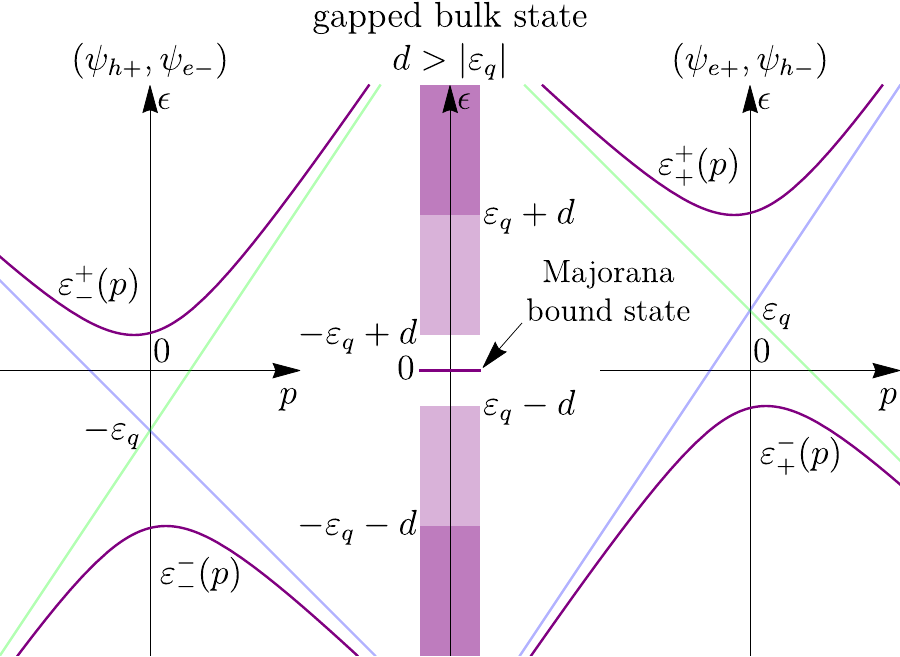}
\caption{
Ever-present Majorana bound state in the gapped superconducting state
of the low-energy model with 1FS [\eqs{H}{bcNR}], the main result of this work.
Our model includes the effect of the Fermi-point mismatch (\figr{system})
and the one-harmonic coordinate dependence $\De_q(x)=\De_0\ex^{\ix qx}$ [\eq{Deq}]
of the pairing field that can help mitigate it.
The bulk spectrum \eqn{ebulk} has individual energy gaps $(\eps_q-d,\eps_q+d)$ and $(-\eps_q-d,-\eps_q+d)$
around $\pm k_0$ momenta, respectively, that open around the modified mismatch energies $\pm\eps_q$
for any magnitude $\De_0$ of the pairing field [\eqs{eq}{d}].
For large enough $\De_0$, such that $d>|\eps_q|$, the bulk state is gapped since
the individual gaps overlap resulting in the global gap $(-d+|\eps_q|,d-|\eps_q|)$.
We find that for the general normal-reflection BCs \eqn{bcNR},
describing a boundary that is an interface with vacuum (sample termination) or an insulator,
there always exists a nondegenerate Majorana bound state at the Fermi level $\e=0$
in this gapped bulk state.
}
\lbl{fig:gapped}
\end{figure}

\begin{figure}
\includegraphics[width=.45\textwidth]{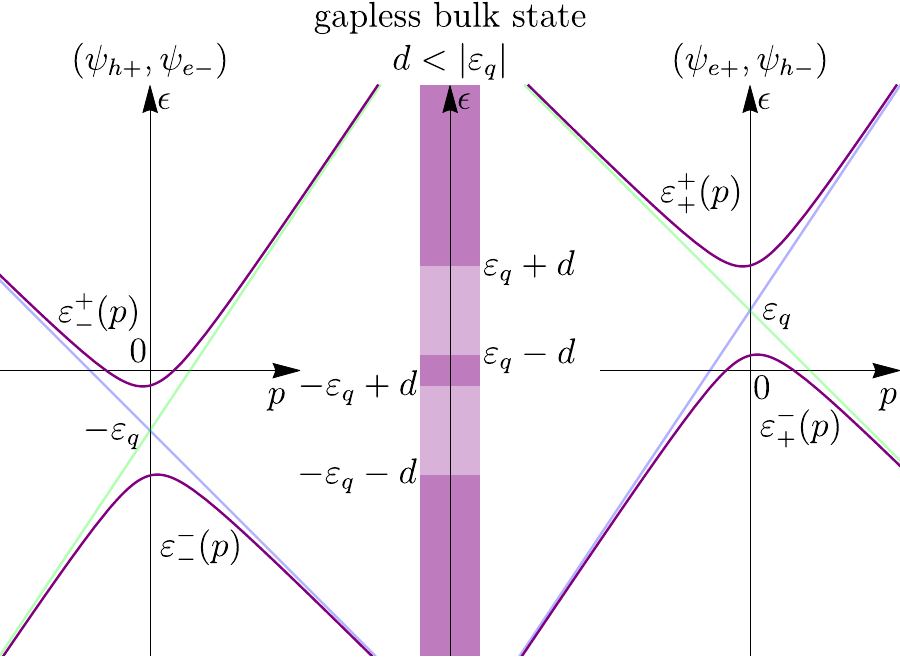}
\caption{Gapless bulk state despite the presence of the superconducting pairing field.
In the presence of the Fermi-point mismatch,
if the amplitude $\De_0$ of the pairing field $\De_q(x)=\De_0\ex^{\ix q x}$ [\eq{Deq}]
is small, such that $d<|\eps_q|$ [\eqs{eq}{d}],
the individual energy gaps $(\eps_q-d,\eps_q+d)$ and $(-\eps_q-d,-\eps_q+d)$
in the bulk spectrum \eqn{ebulk} around $\pm k_0$ points do not overlap and hence, the bulk state is gapless.
Bound states cannot exist in this regime.
}
\lbl{fig:gapless}
\end{figure}

We start by deriving the most general form of the low-energy BdG Hamiltonian for the case of 1FS.

For a real spinful electron system, creating 1FS
requires some spin-orbit {\em or} magnetic effects, {\em or} their combination,
in order to properly split the bands.
An example of a 1FS system with spin-orbit interaction, but absent magnetic effects
is the edge of a 2D quantum spin Hall (QSH) system~\cite{BHZ,Fu}, \figr{qsh}
(anticipating our findings, magnetic effects that break TR symmetry $\Tc_{e-}$
are still required in such system in order to create a boundary).
An example of a 1FS system with only magnetic but no spin-orbit effects
is a quantum wire with a simple quadratic spectrum and Zeeman effect;
inducing superconductivity in it at energies below the Zeeman splitting
is only possible for a spin-triplet pairing field.
In order to induce superconductivity from a spin-singlet pairing field,
spin-orbit interactions are necessary,
which amounts to the proposal of Refs.~\ocite{LdS,Oreg}, \figr{qw}.

Suppose $\Ec_e(k)$ is the normal-state electron band
of the underlying microscopic model
that crosses the Fermi level at two Fermi points
assumed to be at momenta $\pm k_\pm \gtrless 0$ in the 1D Brillouin zone,
$\Ec_e(\pm k_\pm)=0$; all energies will be measured relative to the Fermi level.
Importantly, since we generally assume no symmetries, besides CC,
the Fermi points are not necessarily at opposite momenta;
if $k_+\neq k_-$, we will refer to such situation as the {\em Fermi-point mismatch}, \figr{system}.
The Fermi-point mismatch could be prohibited by a symmetry
(spatial, TR $\Tc_{e\pm}$, or their combination, see, e.g., \appr{T})
that relates $k\rarr -k$ and hence imposes the constraint $\Ec_e(k)=\Ec_e(-k)$.

The corresponding hole spectrum in the normal state (in the absence of superconducting pairing)
is $\Ec_h(k)=-\Ec_e(-k)$ (obtained by applying CC symmetry $\Cc_+$, see below).
The crossings of the electron and hole normal-state spectra
occur at two opposite momenta $\pm k_0$, where
\[
    \Ec_e(\pm k_0)=-\Ec_e(\mp k_0).
\]
For present Fermi-point mismatch, $k_+\neq k_-$,
the crossings occur at finite energies $\pm\eps_0$, respectively.

As we show below, in order for the gapped superconducting state to be induced in such system,
the superconducting pairing field has to overcome the effect of the Fermi-point mismatch.
Therefore, considering the low-energy model,
we assume the mismatch small compared to the nonlinearity scale of the band $\Ec_e(k)$,
i.e., that, at the very least, $|k_+-k_-|\ll k_\pm$.
Under this approximation, the energy and momentum of the crossing points can be found as
\beq
    \eps_0=-\f{v_+v_-}{v_++v_-}(k_+-k_-),\spc
    k_0=\f{v_+k_++v_-k_-}{v_++v_-}.
\lbl{eq:e0}
\eeq
from the linearized electron spectrum
\[
    \Ec_e(k)\approx \pm v_\pm (k\mp k_\pm)=\pm v_\pm(k\mp k_0)\pm\eps_0
\]
around the Fermi or crossing points.
To leading order, the velocities
\[
    v_\pm=\pm\pd_k\Ec_e(\pm k_\pm)\approx \pm\pd_k\Ec_e(\pm k_0)>0
\]
at the Fermi points $\pm k_\pm$ and at the crossing points $\pm k_0$ are equal.

This expansion of the underlying spectrum $\Ec_e(k)$
about the crossing points $\pm k_0$ to the linear order
gives the low-energy electron Hamiltonian
\beq
    \Hh_e(\ph)=
    \lt(\ba{cc}
        v_+\ph+\eps_0 & 0 \\
        0&-v_-\ph-\eps_0
    \ea\rt)
\lbl{eq:He}
\eeq
for the low-energy electron wave function
\beq
    \psih_e(x)=\lt(\ba{c} \psi_{e+}(x) \\ \psi_{e-}(x) \ea\rt).
\lbl{eq:psie}
\eeq
Here, $\ph=-\ix\pd_x$ is the momentum operator
corresponding to the momentum deviations from the crossing points $\pm k_0$.

In the BdG formalism of superconductivity~\cite{BdG},
one introduces a complementary hole wave function
\beq
    \psih_h(x)=\lt(\ba{c} \psi_{h+}(x) \\ \psi_{h-}(x)\ea\rt),
\lbl{eq:psih}
\eeq
so that the full BdG wave function reads
\beq
    \psih(x)=\lt(\ba{c} \psih_e(x)\\ \psih_h(x) \ea\rt)
    =\lt(\ba{c}\psi_{e+}(x)\\\psi_{e-}(x)\\\psi_{h+}(x)\\\psi_{h-}(x)\ea\rt).
\lbl{eq:psi}
\eeq

The structure of the BdG Hamiltonian $\Hh(\ph)$ in this electron-hole space is governed by charge-conjugation (CC) symmetry.
A Hamiltonian $\Hh(\ph)$ satisfies CC symmetry, if there exists an antiunitary CC operation
\beq
    \Cc_+=\Ch_+\Kc,
\lbl{eq:C+}
\eeq
where $\Ch_+$ is a unitary matrix and $\Kc$ is the complex-conjugation operation,
under which the Hamiltonian changes its sign,
\beq
    \Ch_\pm [\Hh(\ph)]^*\Ch_\pm^\dg=-\Hh(\ph),
\lbl{eq:CH}
\eeq
with $\ph^*=-\ph$.
A constraint is also imposed that the CC operation squares to either $\pm 1$.
For a spinful electron system with broken spin symmetry,
the physical CC operation $\Cc_+$ is the one that squares to $+1$
(which we emphasize with the subscript in $\Cc_+$), i.e.,
\[
    \Cc_+^2=\Ch_+\Ch_+^*=\um_4;
\]
it is the one that ensures the antisymmetry of the pairing field,
in accord with the Fermi statistics.
This symmetry alone realizes systems of class D of the topological classification scheme~\cite{Chiu}.

In the basis of \eq{psi}, without loss of generality,
the CC operation can always be brought to the form
\beq
    \Ch_+=\tau_x\ot\um_2
\lbl{eq:Ch+}
\eeq
by the appropriate choice of electron and holes basis states,
where $\tau_x$ is the second Pauli matrix and $\um_2$ is a $2\tm 2$ unit matrix.
So, the action of the CC operation on the wave function reads
\beq
    \Cc_+\psih(x)
    =\lt(\ba{c} \psih_h^*(x) \\ \psih_e^*(x) \ea\rt).
\lbl{eq:C+psi}
\eeq
Note that the CC operation does not alter the coordinate,
which will be important for the analysis of the CC symmetry of the BCs.

Once the CC operation $\Cc_+$ is specified,
the most general form of the BdG Hamiltonian allowed by $\Cc_+$ symmetry reads
\beq
    \Hh(\ph)=
    \lt(\ba{cccc}
        \Hh_e(\ph) & \Deh(x) \\ \Deh^\dg(x) & -\Hh_e^\Tx(-\ph)
    \ea\rt),
\lbl{eq:Hgen}
\eeq
in the basis \eqn{psi}.
The form $-\Hh_e^\Tx(-\ph)$ of the hole Hamiltonian is fixed by that of the electron one [\eq{He}]
and the superconducting pairing field matrix $\Deh(x)$ must be antisymmetric, $\Deh^\Tx(x)=-\Deh(x)$.
Here, $\Tx$ denotes matrix transposition.
And since in the 1FS case it is a $2\tm 2$ matrix,
its form is uniquely fixed by this, $\Deh(x)=\ix\tau_y\De(x)$.
The pairing field of the low-energy model \eqn{Hgen} is therefore fully described by one complex function $\De(x)$.
This holds regardless of the actual underlying spin structure of the pairing field
(this can mean, however, that pairing fields with some spin structures cannot be induced;
see discussion in \secr{qwleH} for the generalized quantum-wire model).

So, for 1FS, the most general form of the
$\Cc_+$-symmetric BdG low-energy Hamiltonian reads
\beq
    \Hh(\ph)=
    \lt(\ba{cccc}
        v_+\ph+\eps_0 & 0 &0&\De(x)\\
        0&-v_-\ph-\eps_0 &-\De(x)&0\\
        0&-\De^*(x)& v_+\ph-\eps_0&0\\
        \De^*(x)&0&0&-v_-\ph+\eps_0
    \ea\rt).
\lbl{eq:H}
\eeq
The pairs $(\psi_{e+}(x),\psi_{h-}(x))$ and $(\psi_{h+}(x),\psi_{e-}(x))$
of the wave-function \eqn{psi} components of the states around $\pm k_0$ points, respectively,
are not coupled by the bulk Hamiltonian \eqn{H}.

The normal-state spectrum in the absence of the pairing field $\De(x)$
for the full BdG Hamiltonian \eqn{H}, including the hole states, is shown in \figr{system}.
Note that both the electron $\psi_{e\pm}(x)$ and hole $\psi_{h\pm}(x)$
components of the BdG wave function \eqn{psi}
are labeled with $\pm$ according to their propagation direction
(right and left, respectively, in the geometry of \figr{system})
as set by the signs $\pm v_\pm\gtrless0$ of their velocities.
The electron components $\psi_{e\pm}(x)$ correspond to $\pm k_0$ momenta, respectively, by construction,
while the hole components $\psi_{h\pm}(x)$ correspond to $\mp k_0$.

\section{General charge-conjugation-symmetric boundary conditions \label{sec:bc}}

\begin{figure}
\includegraphics[width=.48\textwidth]{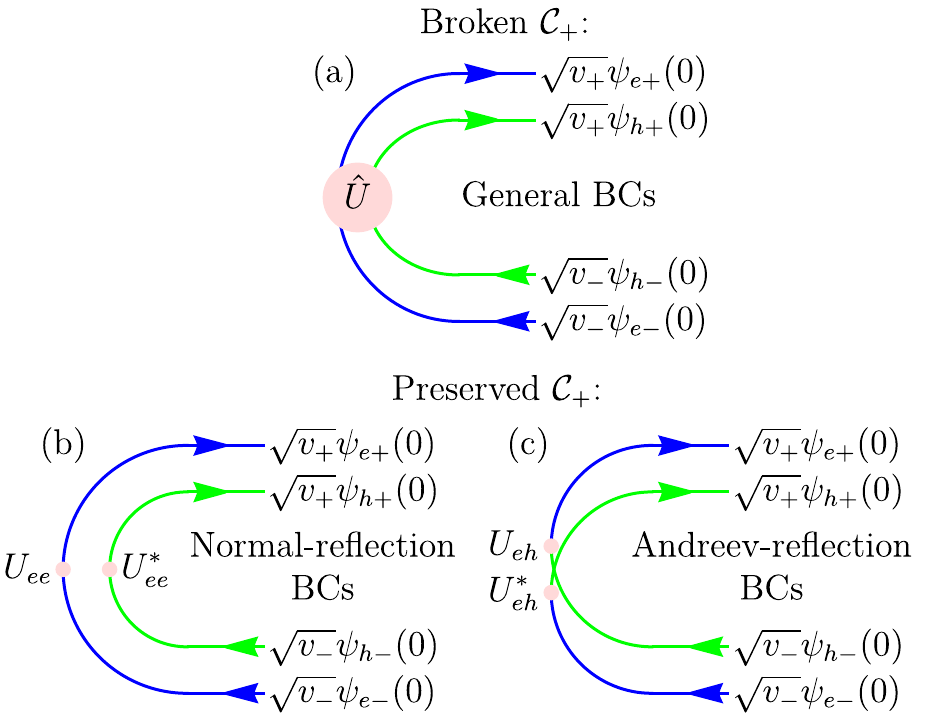}
\caption{
Illustration of the boundary conditions (BCs) for the low-energy model of a 1D superconductor with one Fermi surface (1FS)
with the Hamiltonian $\Hh(\ph)$ [\eq{H}].
(a) The most general form \eqn{bcgen} of the BCs subject only to the fundamental principle
of probability-current conservation [\eqs{j=0}{j}]
is parameterized by the unitary matrix $\Uh\in\Ux(2)$, which has a natural interpretation of the scattering matrix
between the incident (left-moving) $(\psi_{e-},\psi_{h-})$ and reflected (right-moving) $(\psi_{e+},\psi_{h+})$
components of the BdG wave function \eqn{psi}, \figr{system}.
Under charge-conjugation (CC) symmetry $\Cc_+$ [\eqss{C+}{Ch+}{C+psi}], there are only two allowed disconnected
subfamilies of these BCs, which we term (b) {\em normal-reflection} [\eq{bcNR}] and (c) {\em Andreev-reflection} [\eq{bcAR}] BCs.
(b) In the normal-reflection BCs, electrons are reflected as electrons and holes as holes.
The boundary of a superconductor with an insulator, of interest in this work,
can only be described by normal-reflection BCs.
(c) In the Andreev-reflection BCs, electrons are completely reflected as holes
and vice versa. Such BCs cannot be defined in the normal state and cannot represent such boundary.
}
\lbl{fig:bc}
\end{figure}

We now derive the most general form of the BCs for the BdG Hamiltonian \eqn{H} for 1FS,
restricted only by the probability-current conservation principle and CC symmetry $\Cc_+$.

We assume that the low-energy system is half-infinite and occupies the region $x\geq 0$;
$x=0$ is its boundary. The low-energy wave function $\psih(x)$ [\eq{psi}] is not defined in the region $x<0$;
physically, this means that there is a large excitation gap in the region $x<0$ in the underlying microscopic model,
which renders it inaccessible for the low-energy excitations of $\psih(x)$.

The most general possible form of the BCs of a continuum model
is governed~\cite{Berry,Bonneau,Tokatly,McCann,AkhmerovPRL,AkhmerovPRB,Ostaay,Ahari,KharitonovLSM,KharitonovQAH,Seradjeh}
by the fundamental principle of the conservation of the probability current $j(x)$,
which follows from the hermiticity of the Hamiltonian, which, in turn,
follows from the norm conservation of the wave function.
For a half-infinite system, this principle takes the form of the current nullification at the boundary,
\beq
    j(0)=0.
\lbl{eq:j=0}
\eeq
For an arbitrary 1D continuum model, such BCs have been derived in Ref.~\ocite{Ahari}.
For the linear-in-momentum Hamiltonian \eqn{H}, the probability current
\beq
    j(x)=j_e(x)+j_h(x)
    ,\spc
    j_\nu(x)=j_{\nu+}(x)+j_{\nu-}(x),
\lbl{eq:j}
\eeq
\beq
    j_{\nu\pm}(x)=\pm v_\pm \psi_{\nu\pm}^*(x)\psi_{\nu\pm}(x),\spc \nu=e,h,
\lbl{eq:jnupm}
\eeq
is a diagonal quadratic form of the wave-function components and
the most general form of the BCs nullifying it reads
\beq
    \psirh_+(0)=\Uh\psirh_-(0).
\lbl{eq:bcgen}
\eeq
Here,
\[
    \psirh_\pm(x)
    =\lt(\ba{c} \sq{v_\pm} \psi_{e\pm}(x) \\ \sq{v_\pm} \psi_{h\pm}(x) \ea\rt)
\]
are the vectors that group the wave-function components \eqn{psi}
with the same sign $\pm$ of velocities (right- and left-moving states, respectively) and
\[
    \Uh=\lt(\ba{cc} u_{ee} & u_{eh} \\ u_{he} & u_{hh} \ea\rt)
,\spc    \Uh\Uh^\dg=\um,
\]
is a U(2) unitary matrix.
Thus, all possible BCs form a family parameterized by the unitary matrix $\Uh$.
Each instance of $\Uh$ delivers one possible set of BCs.
The structure of the BCs \eqn{bcgen} is particularly transparent
and has a natural physical interpretation
%~\cite{KharitonovBCs}
as a scattering process off the boundary between the incident (left-moving) $\psirh_-(x)$
and reflected (right-moving) $\psirh_+(x)$ waves,
where $\Uh$ plays the role of the scattering matrix.

For arbitrary $\Uh$, these BCs generally break $\Cc_+$ symmetry,
in which case the system with a boundary
does not have $\Cc_+$ symmetry even though the bulk Hamiltonian \eqn{H} does.
However, in order for the bound states to represent the topological properties of the bulk
stemming from a certain symmetry, the system with a boundary must satisfy that symmetry.
Within the continuum-model formalism,
this means that the BCs must also satisfy that symmetry~\cite{KharitonovLSM}.
In turn, the latter means that the wave function
transformed under that symmetry operation also satisfies the BCs
(BCs essentially restrict the Hilbert space, and since the transformed wave function must also
belong to the same Hilbert space, it must satisfy those BCs).
This introduces constraints on the allowed form of $\Uh$.
For the CC symmetry $\Cc_+$ in question, inserting the transformed wave function $\Cc_+\psih(0)$ [\eq{C+psi}]
into the BCs \eqn{bcgen},
we find that the transformed wave function satisfies them if
the following constraints on $\Uh$ are satisfied:
\[
    u_{hh}=u_{ee}^*,\spc u_{he}=u_{eh}^*.
\]
(Note that, importantly, the CC operation $\Cc_+$ does not alter the coordinate $x$
and thus leaves the geometry of system intact;
hence, the system with a boundary can, in principle, be $\Cc_+$-symmetric.)
From this, choosing $u_{ee}$ and $u_{eh}$ as the independent matrix elements
and combining with the unitarity of $\Uh$, we obtain the following constraints:
\[
    u_{ee}u_{ee}^*+u_{eh}u_{eh}^*=1,\spc u_{ee}u_{eh}=0.
\]
Thus, there are only two options.
The first option is when $u_{eh}=0$, and
\[
    \Uh=\lt(\ba{cc} U_{ee} &0\\0&U_{ee}^*\ea\rt)
    ,\spc
    |U_{ee}|=1;
\]
the spelled out \eq{bcgen} takes the form
\beqar
    \sq{v_+}\psi_{e+}(0)&=&U_{ee}\sq{v_-}\psi_{e-}(0), \nn\\
    \sq{v_+}\psi_{h+}(0)&=&U_{ee}^*\sq{v_-}\psi_{h-}(0).
\lbl{eq:bcNR}
\eeqar
The second option is when $u_{ee}=0$,
and
\[
    \Uh=\lt(\ba{cc} 0&U_{eh}\\ U_{eh}^* & 0\ea\rt), \spc
    |U_{eh}|=1;
\]
the spelled out \eq{bcgen} takes the form
\beqar
    \sq{v_+}\psi_{e+}(0)&=&U_{eh}\sq{v_-}\psi_{h-}(0),\spc \nn\\
    \sq{v_+}\psi_{h+}(0)&=&U_{eh}^*\sq{v_-}\psi_{e-}(0).
\lbl{eq:bcAR}
\eeqar

Thus, we obtain that the most general BCs
subject only to the current nullification principle and
CC symmetry $\Cc_+$ consist of two families, \eqs{bcNR}{bcAR},
each parameterized by the respective scattering phase factors $U_{ee}$ and $U_{eh}$;
each value of $U_{ee}$ or $U_{eh}$ corresponds to one possible set of BCs of the respective family.
Note that the two families are disconnected in the parameter space of the $\Uh\in\Ux(2)$ matrix manifold
and can never be connected without breaking $\Cc_+$ symmetry.
Also note that, for each family, one of the equations in the BCs
can be obtained from the other by the $\Cc_+$ operation,
underscoring their $\Cc_+$ symmetry.

In the BCs \eqn{bcNR}, there is no scattering between the electron and hole components of the wave function;
the electron $j_e(0)=j_{e+}(0)+j_{e-}(0)=0$ and hole $j_h(0)=j_{h+}(0)+j_{h-}(0)=0$
contributions to the total current \eqn{j} vanish individually.
On the other hand, in the BCs \eqn{bcAR}, there is scattering only between the electron and hole components;
the combinations $j_{e+}(0)+j_{h-}(0)=0$ and $j_{e-}(0)+j_{h+}(0)=0$,
each involving parts of the electron and hole currents, vanish individually.
Accordingly, we term these BCs the {\em normal-reflection} [\eq{bcNR}]
and {\em Andreev-reflection} [\eq{bcAR}] BCs, respectively.
The BCs are illustrated schematically in \figr{bc}.

The main crucial difference between the two families of BCs
is that the normal-reflection BCs \eqn{bcNR}
are well-defined already in the normal state,
i.e., in the absence of the superconducting pairing field $\De(x)$
and just for the electron part $\psih_e(x)$ of the BdG wave function $\psih(x)$ [\eq{psi}],
without introducing the hole part $\psih_h(x)$.
Indeed, the first BC in \eq{bcNR} involves only the components of $\psih_e(x)$ and
is the most general form of the BC~\cite{Berry,AkhmerovPRL,AkhmerovPRB,Ostaay,Ahari,KharitonovLSM,KharitonovQAH}
for the electron Hamiltonian $\Hh_e(\ph)$ [\eq{He}]
with the probability current $j_e(x)$ [\eq{j}]
subject only to the current nullification principle $j_e(0)=0$.

In contrast, the Andreev-reflection BCs \eqn{bcAR} cannot be defined in the normal state:
they can be defined only in the presence of the hole part $\psih_h(x)$,
which implies the presence of superconductivity at the boundary in some form even in the absence of the pairing field $\De(x)$.
For example, one plausible physical realization of the Andreev-reflection BCs
could be that the inaccessible region $x<0$ is a superconductor with a much larger gap.

In this work, our focus is on the bound states of
a physical quasi-1D sample that is terminated (as in \figr{qw}) or interfaced with an insulator (as in \figr{qsh}),
when the inaccessible region $x<0$ is a vacuum or an insulating material.
According to the above considerations, the boundary of such system with a well-defined normal state
can be represented in the low-energy model only by the normal-reflection BCs \eqn{bcNR}.
Therefore, in the remainder of the paper, we will consider only the normal-reflection BCs \eqn{bcNR},
while the Andreev-reflection BCs \eqn{bcAR} will be explored elsewhere.

\section{Fermi-point mismatch, coordinate-dependent pairing field, bulk spectrum \lbl{sec:Deq}}

The derived low-energy Hamiltonian \eqn{H} is valid for an arbitrary coordinate dependence of the pairing field $\De(x)$.
Which dependence is actually favored in a real system may depend on various factors and be nonobvious.
As the main application, we have in mind the setups
where superconductivity is induced in the quasi-1D system due to the proximity effect
of a nearby superconductor, as in \figsr{qw}{qsh}.
Strictly speaking, within the mean-field approach to the interacting many-body Hamiltonian,
the favored form of the pairing field must be found
by minimizing the energy of the BCS Slater-determinant
trial many-body state of this coupled system.
Without the Fermi-point mismatch,
it is likely that the induced pairing field is coordinate-independent, $\De(x)=\De_0$.
With the Fermi-point mismatch, however,
a plausible scenario is that the induced pairing field
could have a periodic behavior associated with the momentum mismatch $k_+-k_-\neq 0$,
akin to the Larkin-Ovchinnikov-Fulde-Ferrel state~\cite{LO,FF}.

Let us examine a one-harmonic periodic coordinate dependence
\beq
    \De_q(x)=\De_0\ex^{\ix qx}, \spc \De_0=|\De_0|\ex^{-\ix\de},
\lbl{eq:Deq}
\eeq
characterized by some momentum $q$, which we discuss below.
This explicit coordinate dependence can be eliminated from
the Hamiltonian $\Hh(\ph)|_{\De(x)=\De_q(x);\eps_0}$ [\eq{H}]
with the pairing field $\De_q(x)$ and Fermi-point-mismatch energy $\eps_0$
by the following transformation of the wave function:
\beq
    \psih(x)=
    \lt(\ba{c}
        \ex^{\ix(+1-\vr_0)\f{q}2x}\psi_{e+}'(x)\\
        \ex^{\ix(+1+\vr_0)\f{q}2x}\psi_{e-}'(x)\\
        \ex^{\ix(-1+\vr_0)\f{q}2x}\psi_{h+}'(x)\\
        \ex^{\ix(-1-\vr_0)\f{q}2x}\psi_{h-}'(x)
    \ea\rt)
    ,\spc
    \psih'(x)
    =\lt(\ba{c}
        \psi_{e+}'(x)\\
        \psi_{e-}'(x)\\
        \psi_{h+}'(x)\\
        \psi_{h-}'(x)
    \ea\rt),
\lbl{eq:psipsi'}
\eeq
where
\beq
    \vr_0=\f{v_0}{v_z}, \spc
     v_{0,z}=\tf12(v_+\mp v_-).
\lbl{eq:v0z}
\eeq
It essentially amounts to introducing proper individual momentum shifts for the wave-function components,
which can be deduced by analyzing the Hamiltonian in momentum space.
This way, we find that the Hamiltonian for the wave function $\psih'(x)$ has the form
\beq
    \Hh'(\ph)\equiv\Hh(\ph)|_{\De(x)=\De_0; \eps_0\rarr\eps_q}
\lbl{eq:H'}
\eeq
of \eq{H}, but with the coordinate-independent pairing field $\De_0$
and modified, $q$-dependent mismatch energy [\eq{e0}]
\beq
    \eps_q=\eps_0+\f{v_+v_-}{v_++v_-}q=\f{v_+v_-}{v_++v_-}[q-(k_+-k_-)].
\lbl{eq:eq}
\eeq

The Hamiltonian $\Hh'(\ph)$ is thus effectively translationally invariant
and its bulk eigenstates for an infinite system
can be characterized by the momentum quantum number $p$
[although different components of the original wave function $\psih(x)$ have different momenta, as per \eq{psipsi'}].
Hence, the bulk spectrum of both Hamiltonians $\Hh'(\ph)$ and $\Hh(\ph)|_{\De(x)=\De_q(x);\eps_0}$
consists of the bands
\beqar
    \eps_+^\pm(p)&=&+\eps_q+v_0 p \pm\sq{v_z^2p^2+|\De_0|^2}, \nn\\
    \eps_-^\pm(p)&=&-\eps_q+v_0 p \pm\sq{v_z^2p^2+|\De_0|^2}
\lbl{eq:ebulk}
\eeqar
for the decoupled pairs $(\psi_{e+}(x),\psi_{h-}(x))$ and $(\psi_{h+}(x),\psi_{e-}(x))$ of components
around $\pm k_0$ momenta, respectively, as plotted in \figsr{gapped}{gapless}.
We observe that the individual energy gaps $(\eps_q-d,\eps_q+d)$ and $(-\eps_q-d,-\eps_q+d)$
with
\beq
    d=|\De_0|\sq{1-\vr_0^2}
\lbl{eq:d}
\eeq
open up in the bulk spectrum \eqn{ebulk} around $\pm k_0$ momenta, respectively,
for any value of the pairing field $\De_0$.
However, when $d<|\eps_q|$, these individual gaps do not overlap and
the bulk state remains gapless despite the presence of the pairing field (\figr{gapless}).
Only when $d>|\eps_q|$ exceeds the modified mismatch energy,
the individual gaps overlap, resulting in the global energy gap $(-d+|\eps_q|,d-|\eps_q|)$
centered around the Fermi level $\e=0$ (\figr{gapped}).
Thus, whenever the modified mismatch energy $\eps_q\neq0$ is nonzero,
there is a threshold for creating a gapped bulk superconducting state~\cite{Nesterov,Rex}.

The modified mismatch energy $\eps_q=0$ is absent when $q=k_+-k_-$.
Only in this case there is no threshold for gap opening,
which would suggest that such $q$ in \eq{Deq} should be favorable.
However, this means that in the nearby superconductor (in setups such as in \figr{qw}),
which is the source of superconductivity,
the pairing field must have a similar coordinate dependence
at least in some region close to the quasi-1D system,
which would cause a penalty in the gradient energy.
Therefore, which value of $q$
(whether $0$ or $k_+-k_-$, or some intermediate value)
minimizes the ground-state energy of the interacting many-body system
cannot be answered without carrying out the minimization procedure for the whole coupled system.
This question is beyond the focus of this work and we do not attempt to answer it here.
Instead, we calculate the bound states for any Fermi-point mismatch $k_+-k_-$
and for the coordinate dependence \eqn{Deq} of the pairing field with any $q$
and demonstrate that a Majorana bound state does exist regardless
of their values, as long as the superconducting state is gapped (\figr{gapped}).

\section{Ever-present Majorana bound state for one Fermi surface \lbl{sec:bs}}

Having derived the general forms of the bulk Hamiltonian [\eq{H}]
and normal-reflection BCs [\eq{bcNR}] for the case of 1FS,
we now analytically calculate the bound states for the half-infinite system $x\geq0$
with the pairing field of the form $\De_q(x)$ [\eq{Deq}].

The calculation is straightforward.
We first construct a general solution to the Schr\"odinger equation
\[
    \Hh'(\ph)\psih'(x)=\e\psih'(x)
\]
that decays into the bulk, as $x\rarr+\iy$.
A decaying solution can exist only in the gapped superconducting state (\figr{gapped}),
i.e., for $d>|\eps_q|$ [\eqs{eq}{d}] in the presence of the Fermi-point mismatch,
at energies $\e\in(-d+|\eps_q|,d-|\eps_q|)$
within the gap of the bulk spectrum \eqn{ebulk}.
The general decaying solution is a linear combination of particular decaying solutions with complex momenta $p(\e)$,
obtained from the characteristic equation
\[
    \x{det}[\Hh'(p)-\e\um_4]=0.
\]
There are two such particular solutions
$\hat{\Xc}_\pm(\e)\ex^{\ix p_\pm(\e) x}$,
with the momenta
\beq
    p_\pm(\e)
    =\f{d}{v_z^2-v_0^2} (-\vr_0\er_\pm+\ix \sq{1-\er_\pm^2})
\lbl{eq:p}
\eeq
and vectors
\beq
    \hat{\Xc}_+(\e)
    =\lt(\ba{c} \tf1{\sq{v_+}}X_{e+}(\e)\\0\\0\\\tf1{\sq{v_-}}\ea\rt)
,\spc
    \hat{\Xc}_-(\e)
    =\lt(\ba{c} 0\\\tf1{\sq{v_-}}\\\tf1{\sq{v_+}}X_{h+}(\e)\\0\ea\rt),
\lbl{eq:Xc}
\eeq
where we denote
\beq
    X_{e+}(\e)=\ex^{\ix[\Phi_+(\e)-\de]},\spc X_{h+}(\e)=-\ex^{\ix[\Phi_-(\e)+\de]},
\lbl{eq:X}
\eeq
\[
    \ex^{\ix\Phi_\pm(\e)}
    =\er_\pm+\ix\sq{1-{\er_\pm}^2}
    ,\spc \er_\pm=\f{\e\mp\eps_q}{d}.
\]
We observe that the relation
\beq
    X_{h+}(\e)=X_{e+}^*(-\e)
\lbl{eq:Xrel}
\eeq
holds, which ultimately is a consequence of the CC symmetry $\Cc_+$
and is key to the conclusion about the Majorana bound state at $\e=0$.

These particular solutions originate from the pairs $(\psi_{e+}'(x),\psi_{h-}'(x))$ and $(\psi_{h+}'(x),\psi_{e-}'(x))$
of components around $\pm k_0$ momenta, respectively, which are decoupled in the bulk Hamiltonian.
The general decaying solution reads
\beq
    \psih'(x)
    =c_+\hat{\Xc}_+(\e)\ex^{\ix p_+(\e) x}
    +c_-\hat{\Xc}_-(\e)\ex^{\ix p_-(\e) x},
\lbl{eq:psi'solution}
\eeq
where $c_\pm$ are the free coefficients.

Inserting this wave function \eqn{psi'solution} via \eq{psipsi'}
into the normal-reflection BCs \eqn{bcNR}, we obtain a linear homogeneous system
\[
    X_{e+}(\e)c_+=U_{ee} c_-
,\spc
    X_{h+}(\e)c_-=U_{ee}^* c_+
\]
of equations for the unknown coefficients $c_\pm$, in which the energy $\e$ is a parameter.
This system has nontrivial solutions, which are the sought bound states, if its determinant is zero,
i.e., when
\[
    X_{e+}(\e)X_{h+}(\e)=U_{ee} U_{ee}^*.
\]
The scattering phase factor $U_{ee}$
drops out due to $U_{ee}U_{ee}^*=1$;
so does the phase $\de$ of the pairing field \eqn{Deq} contained in \eq{X}.
Using the key relation \eqn{Xrel}, the equation
for the energy $\e$ of possible bound states becomes
\beq
    X_{e+}(\e)X_{e+}^*(-\e)=1.
\lbl{eq:Xeq}
\eeq
In this form, it is evident that $\e=0$ is a solution, which describes a Majorana bound state.
In \figr{argXX}, we plot the argument of the left-hand side of \eq{Xeq},
which also shows that there are no other bound-state solutions.

\begin{figure}
\includegraphics[width=.42\textwidth]{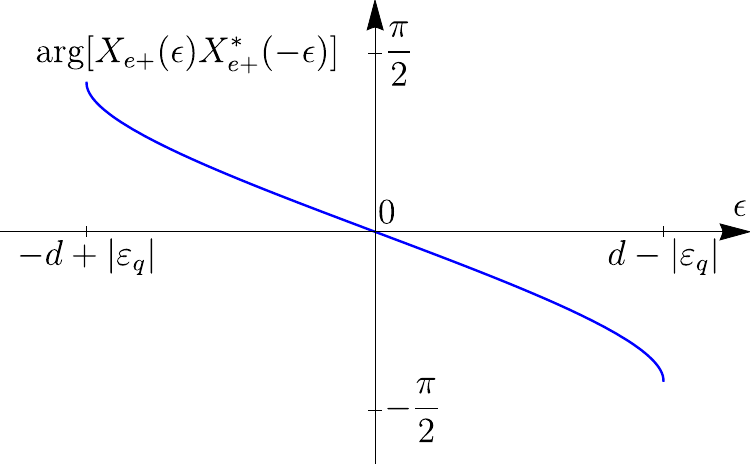}
\caption{
The argument of the left-hand side $X_{e+}(\e)X_{e+}^*(-\e)$ [\eq{X}]
of \eq{Xeq} determining the energies of possible bound states,
as a function of energy $\e\in(-d+|\eps_q|,d-|\eps_q|)$ within the gap.
This plot explicitly shows that there exists a Majorana bound state at $\e=0$
and there are no other bound-state solutions.
%$\eps_q/d=0.628074$
}
\lbl{fig:argXX}
\end{figure}

Hence, we arrive at the central result of this work, that,
as long as a gapped superconducting state can be induced in a quasi-1D system with 1FS,
there always exists a Majorana bound state at its boundary with an insulator
(where the latter could be an actual insulator material or vacuum, for a terminated sample),
as illustrated in \figr{gapped}.
As already explained in \secr{intro},
since this claim has been proven for the low-energy model
with the most general forms of the bulk Hamiltonian and normal-reflection BCs,
it holds for the whole class of systems:
any microscopic model with such general properties
will reduce to an instance of the derived general low-energy model in the low-energy limit
and will thus host Majorana bound states regardless of its other details.
Although the presented proof is self-contained,
it is nonetheless insightful to illustrate
this latter point for specific microscopic models,
which we do in \secsr{qw}{qsh}.

We stress the $\e=0$ Majorana bound state
exists specifically for normal-reflection BCs \eqn{bcNR},
and only this family of general $\Cc_+$-symmetric BCs
can represent the boundary with an insulator, which is the focus of the present work.
As explained in \secr{bc}, Andreev-reflection BCs cannot represent such boundary,
since they imply the presence of superconductivity in some form even for absent pairing field $\De(x)$.
The analysis of Andreev-reflection BCs is for this reason beyond the focus of this work.

\section{
Ever-present Majorana bound state for odd number of Fermi surfaces
\lbl{sec:NFS}}

In this section, we generalize the above-presented formalism and result
to the case of arbitrary number $N\geq 1$ of FSs.
The wave function for the low-energy BdG model now reads
\[
    \psih(x)=\lt(\ba{c} \psih_{e+}(x) \\ \psih_{e-}(x) \\ \psih_{h+}(x) \\ \psih_{h-}(x) \ea\rt)
\]
where
\[
    \psih_{\nu\pm}(x)=\lt(\ba{c} \psi_{\nu\pm}^1 (x) \\ \ldots \\ \psi_{\nu\pm}^N (x) \ea\rt),\spc \nu=e,h,
\]
are now vectors in the $N$-dimensional FS space.
The matrix of the CC symmetry operation now reads
\beq
    \Ch_+=\tau_x\ot\um_2\ot\um_N.
\lbl{eq:C+N}
\eeq

We will consider the case when the separations
between the FSs are much larger than the low-energy scales,
set by the superconducting pairing field and the Fermi-point mismatch
of each FS.
In this case, potential superconducting pairing between different FSs is energetically unfavorable and we neglect it.
Accordingly, the general Hamiltonian
\beq
    \Hh(\ph)=
    \lt(\ba{cccc}
        \vh_+\ph+\epsh_0 &\nm&\nm&\Deh(x)\\
        \nm&-\vh_-\ph-\epsh_0 &-\Deh(x)&\nm\\
        \nm&-\Deh^\dg(x)& \vh_+\ph-\epsh_0&\nm\\
        \Deh^\dg(x)&\nm&\nm&-\vh_-\ph+\epsh_0
    \ea\rt)
\lbl{eq:HN}
\eeq
is diagonal in the FS space and
has the same structure in the Gor'kov-Nambu space as the Hamiltonian \eqn{H} for 1FS.
The velocities
\[
    \vh_\pm=\x{diag}(v_\pm^1,\ldots,v_\pm^N), \spc v_\pm^n>0,
\]
energy shift
\[
    \epsh_0=\x{diag}(\eps_0^1,\ldots,\eps_0^N),
\]
and pairing field
\[
    \Deh(x)=\x{diag}(\De^1(x),\ldots,\De^N(x))
\]
are now diagonal matrices in the FS space.

Turning to the BCs,
the expressions for the probability current \eqn{j} now read
\[
    j_{\nu\pm}(x)=\pm\psih^\dg_{\nu\pm}(x)\vh_\pm\psih_{\nu\pm}(x)=\pm\psirh^\dg_{\nu\pm}(x)\psirh_{\nu\pm}(x),
\]
where
\beq
    \psirh_\pm(x)=\lt(\ba{c} \psirh_{e\pm}(x) \\ \psirh_{h\pm}(x)\ea\rt)
    ,\spc
    \psirh_{\nu\pm}(x)=\lt(\ba{c} \sq{v_\pm^1}\psi_{\nu\pm}^1(x) \\\ldots\\ \sq{v_\pm^N}\psi_{\nu\pm}^N(x) \ea\rt).
\lbl{eq:psir}
\eeq
According to Ref.~\ocite{Ahari},
the BCs of the most general form, satisfying only the current-nullification principle, have the form
\beq
    \psirh_+(0)=\Uh\psirh_-(0),
\lbl{eq:bcgenN}
\eeq
with $\psirh_\pm(x)$ from \eq{psir} and an arbitrary unitary matrix
\beq
    \Uh=\lt(\ba{cc} \uh_{ee} & \uh_{eh} \\ \uh_{he} & \uh_{hh} \ea\rt), \spc \Uh\Uh^\dg=\um_{2N}.
\lbl{eq:UN}
\eeq
Applying CC symmetry $\Cc_+$ [\eq{C+N}] to these BCs gives the relations
\beq
    \uh_{he}=\uh_{eh}^*,\spc \uh_{hh}=\uh_{ee}^*.
\lbl{eq:u*N}
\eeq
Choosing the matrices $\uh_{ee}$ and $\uh_{eh}$ as independent parameters,
the unitarity of $\Uh$ reduces to the following constraints:
\beq
    \uh_{ee}\uh_{ee}^\dg+\uh_{eh}\uh_{eh}^\dg=\um_N,\spc
    \uh_{ee}\uh_{eh}^\Tx+\uh_{eh}\uh_{ee}^\Tx=\nm_N.
\lbl{eq:ucN}
\eeq

The equation \eqn{bcgenN} with \eq{psir} and the unitary matrix \eqn{UN} satisfying the constraints \eqsn{u*N}{ucN}
is thus the most general form of the BCs of a $\Cc_+$-symmetric system with a boundary,
which describe all possible interfaces with all possible microscopic structures,
and where the inaccessible region $x<0$ could be an insulator or a superconductor.

Explicitly resolving the constraints \eqn{ucN} for $N>1$ is not trivial and we do not attempt it here.
Instead, we derive a subset of the general BCs, which we term {\em normal-reflection} BCs,
that describe the boundary  between a superconductor and an insulator, of interest in this work.
For such an interface, there cannot be superconducting coupling between electrons and holes originating from it.
Hence, the matrix $\uh_{eh}=\nm$ responsible for such processes must vanish.
The remaining matrix $\uh_{ee}=\Uh_{ee}$ of order $N$ is then a unitary matrix [\eq{ucN}],
describing scattering within the electron subspace only.
The full scattering matrix is block-diagonal:
\beq
    \Uh=\lt(\ba{cc} \Uh_{ee} & \nm \\ \nm & \Uh_{ee}^* \ea\rt), \spc \Uh_{ee}\Uh_{ee}^\dg=\um_N.
\lbl{eq:UNNR}
\eeq
These are the normal-reflection BCs of the most general form, describing an arbitrary interface with an insulator.
Note that for such normal-reflection BCs, the electron $j_{e+}(0)+j_{e-}(0)=0$
and hole $j_{h+}(0)+j_{h-}(0)=0$ currents vanish independently.

Next, we analyze the bound states.
As in the 1FS case (\secr{Deq}), we consider one-harmonic coordinate dependencies
\beq
    \De^n_q(x)=\De_0^n\ex^{\ix q^n x},
    \spc
    \De_0^n=|\De_0^n|\ex^{-\ix\de^n},
\lbl{eq:Deqn}
\eeq
with independent momenta $q^n$ of the pairing fields within each FS;
such forms could help mitigate the effect of the Fermi-point mismatch.
These explicit coordinate dependencies can be eliminated
via the change of basis from $\psih(x)$ to $\psih'(x)$ and $\Hh(\ph)$ to $\Hh'(\ph)$,
described by the same Eqs.~\eqn{psipsi'}-\eqn{eq} within each FS subspace.

Again, we look for the general solution to the Schr\"odinger equation
\[
    \Hh'(\ph)\psih'(x)=\e\psih'(x)
\]
that decays into the bulk.
Since the FSs are not coupled by the Hamiltonian,
there are independent particular solutions $\Xch_\pm^n(\e)\ex^{\ix p_\pm^n(\e)x}$ for each FS $n$,
with momentum solutions $p_\pm^n(\e)$ [\eqn{p}] to the characteristic equation
\[
    \x{det}[\Hh'(p)-\e\um_{4N}]=0
\]
and eigenvectors $\Xch_+^n(\e)$, which have the structure of \eq{Xc} in the Gor'kov-Nambu subspace of the $n$-th FS
and are zero in the subspaces of other FSs.
The general decaying solution
\beq
    \psih'(x)
    =\sum_{n=1}^N[c_+^n\Xch_+^n(\e)\ex^{\ix p_+^n(\e) x}
    +c_-^n\Xch_-^n(\e)\ex^{\ix p_-^n(\e) x}]
\lbl{eq:psiN'solution}
\eeq
is their linear combination with the free coefficients $c_\pm^n$.

Inserting this general solution $\psih(0)=\psih'(0)$ into the normal-reflection BCs [\eqs{bcgenN}{UNNR}],
we obtain the equations
\beq
    \Xh_{e+}(\e)\ch_+
    =\Uh_{ee}\ch_-
    ,\spc
    \Xh_{h+}(\e)\ch_-
    =\Uh_{ee}^*\ch_+
\lbl{eq:Xceq}
\eeq
(note that velocity factors completely drop out again),
where
\[
    \ch_\pm=\lt(\ba{c} c_\pm^1 \\\ldots\\ c_\pm^N \ea\rt)
\]
are the vectors of the free coefficients and
\[
    \Xh_{\nu+}(\e)=\x{diag}(X_{\nu+}^1(\e),\ldots,X_{\nu+}^N(\e))
\]
are diagonal matrices in the FS space, with $X_{\nu+}^n(\e)$ being the factors \eqn{X} for each FS.
From the scalar relation \eqn{Xrel} for one 1FS, the analogous relation follows
for the diagonal matrices:
\beq
    \Xh_{h+}(\e)=\Xh_{e+}^*(-\e).
\lbl{eq:Xhrel}
\eeq

Excluding $\ch_+$ from \eq{Xceq} and using the unitarity of $\Uh_{ee}$,
we arrive at the equation
\[
    \Wh(\e)\ch_-=\nm
\]
for $\ch_-$, with
\[
    \Wh(\e)=\Uh_{ee}-\Xh_{e+}(\e)\Uh_{ee}^\Tx\Xh_{e+}^*(-\e).
\]
A nontrivial solution for $\ch_-$ at some energy $\e$ exists, if the matrix $\Wh(\e)$ is degenerate.
Hence, the energies of possible bound states are determined from the equation
\[
    \det\Wh(\e)=0.
\]

Further analysis follows closely that of Ref.~\ocite{Samokhin};
however, as we show, the result holds
for a more general Hamiltonian \eqn{HN} that does not assume $\Tc_\pm$ symmetries,
and thus includes unequal velocities $v_\pm^n$ of the right- and left-movers,
Fermi-point mismatches $\eps_0^n$, and coordinate dependencies \eqn{Deqn} of the pairing fields.

Using the key relation \eqn{Xhrel}
and the obvious relation $\Xh_{e+}(\e)\Xh_{e+}^*(\e)=\um$,
we obtain the relation
\[
    \Wh^\Tx(\e)
    =-\Xh^*_{e+}(-\e) \Wh(-\e)\Xh_{e+}(\e).
\]
Taking the determinant of both sides at zero energy leads to the relation
\[
    \det\Wh(0)
    =\det (-\um_N)\det\Wh(0)=(-1)^N\det\Wh(0).
\]
Hence, for odd number $N$ of FSs, when $(-1)^N=-1$,
the determinant necessarily vanishes at zero energy,
\[
        \det\Wh(0)=0.
\]
This means that there exists at least one zero-energy Majorana bound state,
regardless of the parameters of the Hamiltonian [\eq{HN}]
and of the unitary matrix $\Uh_{ee}$ [\eq{UNNR}] specifying the most general form of normal-reflection BCs.
This proves the main claim in the case of any odd number $N$ of FSs.
Note that for more than 1FS, there may generally be additional bound states
that come in pairs of nonzero energies $\pm\eps$ and are thus not protected by symmetry:
upon varying the parameters, these pairs of bound states may disappear by merging with the bulk states.
Also, at some values of parameters, their energies may turn to zero, $\eps=0$,
which will result in additional degenerate Majorana bound states of accidental nature.

\section{Generalized quantum-wire model \lbl{sec:qw}}

In this and the next sections, we demonstrate how instances
of the general low-energy model [\eqs{H}{bcNR}] for 1FS
derived above purely from symmetry considerations and probability-current conservation principle
arise from ``microscopic'' models~\cite{micro} in the low-energy limit.
We present the systematic procedure of deriving
both the Hamiltonian and BCs of the low-energy model from the underlying microscopic model.

As the first example, we consider a model of a quantum wire,
which has a few generalizations relative to that of Refs.~\ocite{LdS,Oreg}:
an arbitrary direction of the Zeeman field
and a general spin structure of the superconducting pairing field.

\subsection{Electron Hamiltonian and its bulk spectrum}

We consider a quantum-wire model for spinful electrons with the wave function
\beq
    \Psih_e(x)=\lt(\ba{c} \Psi_{e\ua}(x) \\ \Psi_{e\da}(x) \ea\rt)
\lbl{eq:qwPsie}
\eeq
described by the following electron Hamiltonian:
\beq
    \Hch_e(\kh)=\Hch_{e0}(\kh)-h_z\sig_z,
\lbl{eq:qwHe}
\eeq
\beq
    \Hch_{e0}(\kh)=(\be \kh^2-\mu)\sig_0-\al\kh\sig_z-h_x\sig_x-h_y\sig_y.
\lbl{eq:qwHe0}
\eeq
Here $\kh=-\ix\pd_x$ is the momentum operator,
$\be>0$ is the curvature coefficient of the quadratic term,
$\al>0$
is the velocity coefficient of the linear spin-orbit term,
$\mu$ is the chemical potential, $\sig_0=\um$ is the unity matrix and $\sig_\ga$, $\ga=x,y,z$, are the spin Pauli matrices.

We assume that without the magnetic field the system (also the one with the boundary) possesses
the reflection symmetry $\Sig_z:z\rarr-z$ along the horizontal $z$ axis perpendicular to the wire,
as shown in \figr{qw}. This $z$ axis is also the direction of the
average spin-orbit field $[\lan \Eb(\rb)\ran \tm\kb]$,
where the microscopic electric field $\Eb(\rb)$ is on average oriented
along the vertical $y$ direction due to the broken reflection symmetry along it and assumed $\Sig_z$ symmetry.
Choosing this direction as the $z$ spin quantization axis gives the spin-orbit term $-\al\sig_z\kh$.
Next, we consider the magnetic field $\Bb=(B_x,B_y,B_z)$ of arbitrary orientation.
The total Zeeman term, assumed to have full spin rotational symmetry,
reads $-h_x\sig_x-h_y\sig_y-h_z\sig_z$,
with $\hb=\tf{g}2\mu_B\Bb$, where $\mu_B$ is the Bohr magneton.

As a result, the full electron Hamiltonian $\Hc_e(\kh)$ has axial spin rotation symmetry about the $z$ axis.
Its bulk spectrum [\figr{qw}(b)]
\beq
    \Ec_{e\pm}(k)=\be k^2-\mu\pm\sq{(\al k+h_z)^2+h_\p^2}
\lbl{eq:Ee}
\eeq
depends only on the amplitude $h_\p$, but not the angle $\phi_\p$, of the part
\[
    (h_x,h_y)=h_\p(\cos\phi_\p,\sin\phi_\p)
\]
of the Zeeman field in the vertical $xy$ plane containing the wire.
The $h_z$ Zeeman term causes the Fermi-point mismatch due to the asymmetry $\Ec_{e\pm}(k)\neq\Ec_{e\pm}(-k)$,
which disfavors superconductivity, as discussed in \secr{Deq}.

For absent $h_z=0$, $\Hch_{e0}(\kh)$ is the electron Hamiltonian
of the quantum-wire model introduced in Ref.~\ocite{LdS,Oreg}.
As noticed in Refs.~\ocite{Sprb95,Sann,Samokhin} and as we discuss in \appr{qwT},
in this case, the electron system has an effective TR symmetry $\Tc_{e+}$ with $\Tc_{e+}^2=+1$,
which prohibits the Fermi-point mismatch.
The bulk spectrum of the electron Hamiltonian $\Hch_{e0}(k)$ without the $h_z$ term reads [\figr{qw}(b)]
\beq
    \Ec_{e0\pm}(k)=\be k^2-\mu\pm\sq{\al^2 k^2+h_\p^2}.
\lbl{eq:Ee0}
\eeq
Whenever the $h_\p$ Zeeman field is present, there is a range $\mu\in(-h_\p,h_\p)$
of the chemical potential, where the system has 1FS,
in which a Majorana bound state was originally predicted in Refs.~\ocite{LdS,Oreg}.
This is the regime of interest in this work.
Since we assume the effect of the Fermi-point mismatch small in the low-energy limit,
we will take the $h_z$ Zeeman term into account perturbatively.

There are two regimes of the behavior of the lower band $\Ec_{e0-}(k)$, see \figr{qwphi}: (i) for $h_\p<\eps_\al$, where
\beq
    \eps_\al=\f{\al^2}{2\be}
\lbl{eq:ea}
\eeq
is the crossover energy scale,
at which the spin-orbit $\sim \al k$ and quadratic $\sim \be k^2$ terms become comparable,
the lower band has a maximum at $k=0$ and two minima;
(ii) for $h_\p>\eps_\al$, the lower band has only a minimum at $k=0$.
In either regime $h_\p\gtrless\eps_\al$, for $\mu\in(-h_\p,h_\p)$,
the Fermi level crosses only the lower band $\Ec_{e0-}(k)$ twice, at the Fermi points $k=\pm k_0$,
where the Fermi momentum
\beq
    k_0=\sq{\lt(s+\mu+\eps_\al\rt)/\be}, \spc s=\sq{h_\p^2+2 \mu \eps_\al+\eps_\al^2},
\lbl{eq:qwk0}
\eeq
is obtained from $\Ec_{e0-}(k)=0$.
The Fermi points are at opposite momenta $k=\pm k_0$ in the absence of the $h_z$ Zeeman field,
which is the consequence of the effective TR symmetry $\Tc_{e+}$, as we discuss in \appr{qwT}.

The normalized bulk eigenstates of the Hamiltonian $\Hch_{e0}(k)$
for the lower band $\Ec_{e0-}(k)$ are
\beqar
    \chih_e(k)
    &=&\lt(\ba{c} \chi_{e\ua}(k) \\ \chi_{e\da}(k) \ea\rt)\nn\\
    &=&\f1{\sq{N(k)}}\lt(\ba{c} \ex^{-\f{\ix}2\phi_\p} h_\p\\ \ex^{\f{\ix}2\phi_\p}(\sq{\al^2k^2+h_\p^2}-\al k)  \ea\rt),\lbl{eq:chie}\\
    N(k)
    &=&2\sq{\al^2k^2+h_\p^2}(\sq{\al^2k^2+h_\p^2}-\al k),\nn
\eeqar
which will be used in the next section.

\subsection{General form of the superconducting pairing field}

The general BdG Hamiltonian built from the electron Hamiltonian \eqn{qwHe}
of the quantum wire has the standard block structure
\beq
    \Hch(\kh)=\lt(\ba{cc} \Hch_e(\kh) & \Deh[x,\cd] \\ \Deh^{\dg}[x,\cd] & -\Hch_e^\Tx(-\kh) \ea\rt)
\lbl{eq:qwH}
\eeq
for the BdG wave function
\beq
    \Psih(x)=\lt(\ba{c} \Psih_e(x) \\ \Psih_h(x) \ea\rt), \spc
    \Psih_h(x)=\lt(\ba{c} \Psi_{h\ua}(x) \\ \Psi_{h\da}(x) \ea\rt).
\lbl{eq:qwPsi}
\eeq
The charge-conjugation operation reads
\[
    \Cc_+\Psih(x)=\Ch_+\Psih^*(x),\spc \Ch_+=\tau_x\ot\sig_0.
\]

As another generalization of the model proposed in Refs.~\ocite{LdS,Oreg},
we consider the most general form of the superconducting pairing field,
described by the nonlocal integral operator, acting on the wave function as
\beqar
    \Deh[x,\Psih_h]=\int\dx x' \Deh(x,x')\Psih_h(x')
,\spc\nn\\
    \Deh^\dg[x,\Psih_e]=\int\dx x' \Deh^\dg(x',x)\Psih_e(x').
\lbl{eq:qwDe}
\eeqar
Its kernel
\[
    \Deh(x,x')=\sum_{\ga=0,x,y,z}\De_\ga(x,x')\sig_\ga \ix\sig_y
\]
is a matrix function in the spin space consisting of the singlet $\ga=0$
and triplet $\ga=x,y,z$ contributions.
In Refs.~\ocite{LdS,Oreg}, only the singlet term was considered.

As enforced by CC symmetry, the pairing-field operator is antisymmetric, $\Deh(x,x')=-\Deh^\Tx(x',x)$.
As a result, the scalar kernels $\De_\ga(x,x')$ of the singlet and triplet components
are symmetric, $\De_0(x,x')=\De_0(x',x)$,
and antisymmetric, $\De_{x,y,z}(x,x')=-\De_{x,y,z}(x',x)$, respectively.
For this reason, the kernels of the triplet components  must necessarily be nonlocal to be nonzero.

\subsection{Derivation of the low-energy model \lbl{sec:qwleH}}

We now derive the low-energy model from the above microscopic model
of the quantum wire in the 1FS regime, $\mu\in(-h_\p,h_\p)$,
which describes the limit of energies $\e$ close to the Fermi level.
We assume the $h_z$ Zeeman field and the superconducting pairing field \eqn{qwDe}
small and take them into account perturbatively.

For the derivation of the low-energy model, we consider the BdG wave function [\eq{qwPsi}]
\beqar
    \Psih(x)
        &=&\psi_{e+}(x)\lt(\ba{c}\chih_e(+k_0)\\\nm\ea\rt)\ex^{+\ix k_0 x} \nn\\
        &&+\psi_{e-}(x)\lt(\ba{c} \chih_e(-k_0)\\\nm\ea\rt)\ex^{-\ix k_0 x}\nn\\
        &&+\psi_{h-}(x)\lt(\ba{c}\nm\\\chih_h(+k_0)\ea\rt)\ex^{+\ix k_0 x}\nn\\
        &&+\psi_{h+}(x)\lt(\ba{c}\nm\\\chih_h(-k_0)\ea\rt)\ex^{-\ix k_0 x}
\lbl{eq:Psile}
\eeqar
as a linear combination of the bulk plane-wave
eigenstates at the Fermi level $\e=0$ of the Hamiltonian
\beq
    \Hch_0(\kh)=\lt(\ba{cc} \Hch_{e0}(\kh) & \nm \\ \nm  & -\Hch_{e0}^\Tx(-\kh) \ea\rt)
\lbl{eq:qwH0}
\eeq
with the neglected $h_z$ Zeeman term and pairing field \eqn{qwDe}.
There are four such plane-wave bulk eigenstates.
The first and second ones in \eq{Psile} are pure electron solutions
for the lower band $\Ec_{e0-}(k)$ [\eq{Ee0}] at the Fermi points $\pm k_0$, $\Hch_{e0}(\pm k_0)\chih_e(\pm k_0)=\nm$,
with $\chih_e(\pm k_0)$ given by \eq{chie}.
The third and fourth ones in \eq{Psile} with $\chih_h(\pm k_0)=\chih_e^*(\mp k_0)$
are pure hole solutions at the Fermi points $\pm k_0$,
$-\Hch_{e0}^\Tx(\mp k_0)\chih_h(\pm k_0)=\nm$, and are charge conjugates of the former electron solutions.
The coefficients $\psi_{e\pm}(x)$, $\psi_{h\pm}(x)$ of the linear combination
represent the components of the low-energy BdG wave function $\psih(x)$ [\eq{psi}],
whose slow coordinate dependence compared to the scales set by $\mu$, $h_\p$, $\eps_\al$
is meant to account for the deviation of the energy $\e$ from the Fermi level
and for the presence of the small $h_z$ Zeeman term and pairing field.

Inserting the wave function \eqn{Psile} into the full BdG Hamiltonian $\Hch(\kh)$ [\eq{qwH}]
with the pairing field and $h_z$ Zeeman term
and keeping the terms up to the linear order in momentum $\ph\psih(x)$ acting on the low-energy wave function,
we obtain the low-energy BdG Hamiltonian $\Hh(\ph)$
of the form \eqn{H} for $\psih(x)$, with equal velocities $v_\pm=v$,
\beqar
    v&=&\chih_e^\dg(k_0) (2\be k_0\sig_0-\al\sig_z)\chih_e(k_0)\nn\\
    &=&2\be k_0-\f{\al^2k_0}{\sq{\al^2k_0^2+h_\p^2}},
\lbl{eq:qwv}
\eeqar
the Fermi-point mismatch~\cite{k0} energy [\eq{e0}]
\beq
    \eps_0=-\chih_e^\dg(k_0)h_z\sig_z\chih_e(k_0)
    =-\f{\al k_0}{\sq{\al^2k_0^2+h_\p^2}}h_z
\lbl{eq:qwe0}
\eeq
due to the $h_z$ Zeeman field, and the low-energy pairing field
$\Deh(x)=\ix\tau_y\De(x)$ with
\beqar
    \De(x)&=&\chih_e^\dg(k_0)\Deh(x,k_0)\chih_h(k_0)\lbl{eq:qwDele}\\
    &=&\f{\al k_0}{\sq{\al^2k_0^2+h_\p^2}}\De_0(x,k_0)
    +\De_z(x;k_0)\nn\\
    +\f{\ix h_\p}{\sq{\al^2k_0^2+h_\p^2}}[&-&\De_x(x;k_0)\sin\phi_\p+\De_y(x;k_0)\cos\phi_\p],\nn
\eeqar
expressed in terms of the parameters of the microscopic quantum-wire model.
In \eq{qwDele},
\beq
    \De_\ga(x;k)
    =\int\dx x''\,\ex^{-\ix k x''}\De_\ga(x+\tf12x'',x-\tf12x'')
\lbl{eq:qwDeWigner}
\eeq
are the Wigner transforms of the pairing-field kernels.

We observe that in the limit of the 1FS low-energy model,
all spin components $\De_{0,x,y,z}(x;k_0)$ of the microscopic pairing field \eqn{qwDe}
do reduce to just one unique, antisymmetric form $\Deh(x)=\ix\tau_y\De(x)$
of the low-energy pairing field in \eqs{Hgen}{H},
while the spin-orbit and $h_\p$ Zeeman effects control whether a given component contributes or not.
The spin-singlet component $\De_0(x;k_0)$ contributes to $\De(x)$
only if the spin-orbit effect is present, $\al>0$.
The spin-triplet components $\De_{x,y,z}(x;k_0)$
contribute to $\De(x)$ whether the spin-orbit effect is present or not.
The $\De_{x,y}(x;k_0)$ components contribute only when $h_\p$ is present;
interestingly, one may recognize that the combination $-\De_x(x;k_0)\sin\phi_\p+\De_y(x;k_0)\cos\phi_\p$
in which they enter $\De(x)$ is the projection of the vector $(\De_x(x;k_0),\De_y(x;k_0))$
on the direction in the $xy$ plane perpendicular to the vector $(h_x,h_y)$.
The $h_\p$ Zeeman field is per se not necessary for $\De_{0,z}(x;k_0)$ to contribute to $\De(x)$;
however, without it, the 1FS regime is not possible in this system,
since the bands $\Ec_{e0\pm}(k)$ [\eq{Ee0}] would then have a crossing point at $k=0$.

\subsection{Boundary conditions \lbl{sec:bcqw}}

The normal- and Andreev-reflection BCs \eqsn{bcNR}{bcAR} derived in \secr{bc}
represent the two families of all possible BCs for the general low-energy model
that satisfy only the current-conservation principle and $\Cc_+$ symmetry.
When the underlying microscopic model with its boundary is fully specified,
the corresponding BCs of its low-energy model are an instance of these general BCs.
We now explicitly demonstrate this point
by deriving the corresponding BCs for the low-energy wave function $\psih(x)$ [\eq{psi}] from the quantum wire model.
Variants of such systematic derivation procedure
%~\cite{GeneralBCs}
have already been demonstrated for other low-energy models,
e.g., for graphene lattice~\cite{AkhmerovPRB},
Luttinger semimetal~\cite{KharitonovLSM} and quantum anomalous Hall system~\cite{KharitonovQAH}.
This derivation procedure of specific BCs for the low-energy model from an underlying microscopic model
should not be conflated with the derivation procedure of the general BCs
performed in \secr{bc}, based on the current-conservation principle and symmetries.

We assume that the quantum wire occupies the region $x>0$ (\figr{qw})
and consider the hard-wall BCs~\cite{hwbce}
\beq
    \Psih_e(x=0)=\nm
\lbl{eq:hwbce}
\eeq
for the electron part of the BdG wave function \eqn{qwPsie}, which nullify it at the boundary $x=0$.
The corresponding BCs for the hole part are obtained by CC operation, which in this case leads
to the hard-wall BCs
\beq
    \Psih_h(x=0)=\nm
\lbl{eq:hwbch}
\eeq
as well.

For the derivation of the BCs, to the leading order,
one may consider the Hamiltonian $\Hch_0(\kh)$ [\eq{qwH0}],
with the neglected small pairing field and $h_z$ Zeeman field,
instead of the full microscopic Hamiltonian $\Hch(\kh)$ [\eq{qwH}]
and consider the energy right at the Fermi level $\e=0$.
Once the pairing field is neglected, electron $\Psih_e(x)$ and hole $\Psih_h(x)$ parts of
BdG wave function \eqn{qwPsi} are decoupled and one may consider them separately.
This already guarantees that the resulting low-energy BCs
will have the form of the normal-reflection BCs \eqn{bcNR}.

So, we look for a general solution to the Schr\"odinger equation $\Hch_{e0}(\kh)\Psih_e(x)=\nm$
for the electron part of the wave function
that does not grow into the bulk (as $x\rarr+\iy$).
The general solution is a linear combination of particular solutions
with the coordinate dependence $\propto\ex^{\ix k x}$
with generally complex momenta $k$.
The latter are determined from the characteristic equation
\[
    \x{det}\Hch_{e0}(k)
    =(\be k^2-\mu)^2-\al^2k^2-h_\p^2=0,
\]
which has four solutions.
In the regime $\mu\in(-h_\p,h_\p)$ of 1FS we consider,
two of these solutions $k=\pm k_0$ are, as expected,
the real Fermi-point momenta [\eq{qwk0}],
and the corresponding eigenvectors are the bulk states $\chih_e(\pm k_0)$ [\eq{chie}].
The other two solutions $k=\pm\ix \ka$,
$
    \ka=\sq{\lt(s-\mu-\eps_\al\rt)/\be},
$
are imaginary, with the wave functions $\propto \ex^{\mp\ka x}$ decaying and growing into the bulk, respectively.
The eigenvector of the decaying solution $k=\ix\ka$ reads
\[
    \chih_{e0}
    =\lt(\ba{c} \ex^{-\f{\ix}2\phi_\p} h_\p\\ \ex^{\f{\ix}2\phi_\p}[\eps_\al-s+\ix\sq{2\eps_\al(s-\mu-\eps_\al)}]\ea\rt).
\]
Hence, the general solution that does not grow into the bulk reads
\beqar
    \Psih_e(x)=
        &\psi_{e+}\chih_e(+k_0)\ex^{+\ix k_0 x} +\psi_{e-}\chih_e(-k_0)\ex^{-\ix k_0 x}\nn\\
        &+C\chih_{e0}\ex^{-\ka x}.
\lbl{eq:Psiebcqw}
\eeqar
It consists of two bulk plane-wave states and one decaying solution, entering with
arbitrary coefficients $\psi_{e\pm}$ and $C$.
Inserting this form into the BCs \eqn{hwbce}, we obtain two linear relations
\beqar
    \psi_{e+}\chi_{e\ua}(+k_0)+\psi_{e-}\chi_{e\ua}(-k_0)+C\chi_{e0\ua}&=&0,\nn\\
    \psi_{e+}\chi_{e\da}(+k_0)+\psi_{e-}\chi_{e\da}(-k_0)+C\chi_{e0\da}&=&0\nn
\eeqar
for the three coefficients $\psi_{e+},\psi_{e-},C$.
Excluding $C$, we obtain one relation
\beq
    \psi_{e+}=U_{ee}(\mu,h_\p)\psi_{e-}
\lbl{eq:qwbce}
\eeq
for the two coefficients $\psi_{e+}$ and $\psi_{e-}$, where
\beqar
    U_{ee}(\mu,h_\p)
    &=&\ex^{-\ix\phi_{ee}(\mu,h_\p)}\nn\\
    &=&-\f{\chi_{e\ua}(-k_0)\chi_{e0\da}-\chi_{e\da}(-k_0)\chi_{e0\ua}}
        {\chi_{e\ua}(+k_0)\chi_{e0\da}-\chi_{e\da}(+k_0)\chi_{e0\ua}}.
\lbl{eq:qwUee}
\eeqar
The same relation will hold, to leading order,
for the coordinate-dependent components of the electron part of the low-energy BdG wave function \eqn{psi} at the boundary, $\psi_{e\pm}\rarr\psi_{e\pm}(x=0)$.
Hence, the relation \eqn{qwbce} represents the BC for it
and $U_{ee}(\mu,h_\p)$ is indeed the phase factor therein.
The BC for the hole part $\psih_h(x)$ of the low-energy BdG wave function
can be obtained analogously.
We thus  find that indeed the BCs for the low-energy model derived
from the quantum-wire model with the hard-wall BCs \eqsn{hwbce}{hwbch}
have the form \eqn{bcNR} of the general normal-reflection BCs.
Note that taking the decaying solution $\propto \ex^{-\ka x}$ into account in \eq{Psiebcqw}
is essential for deriving the low-energy BCs.

\begin{figure*}
\includegraphics[width=.80\textwidth]{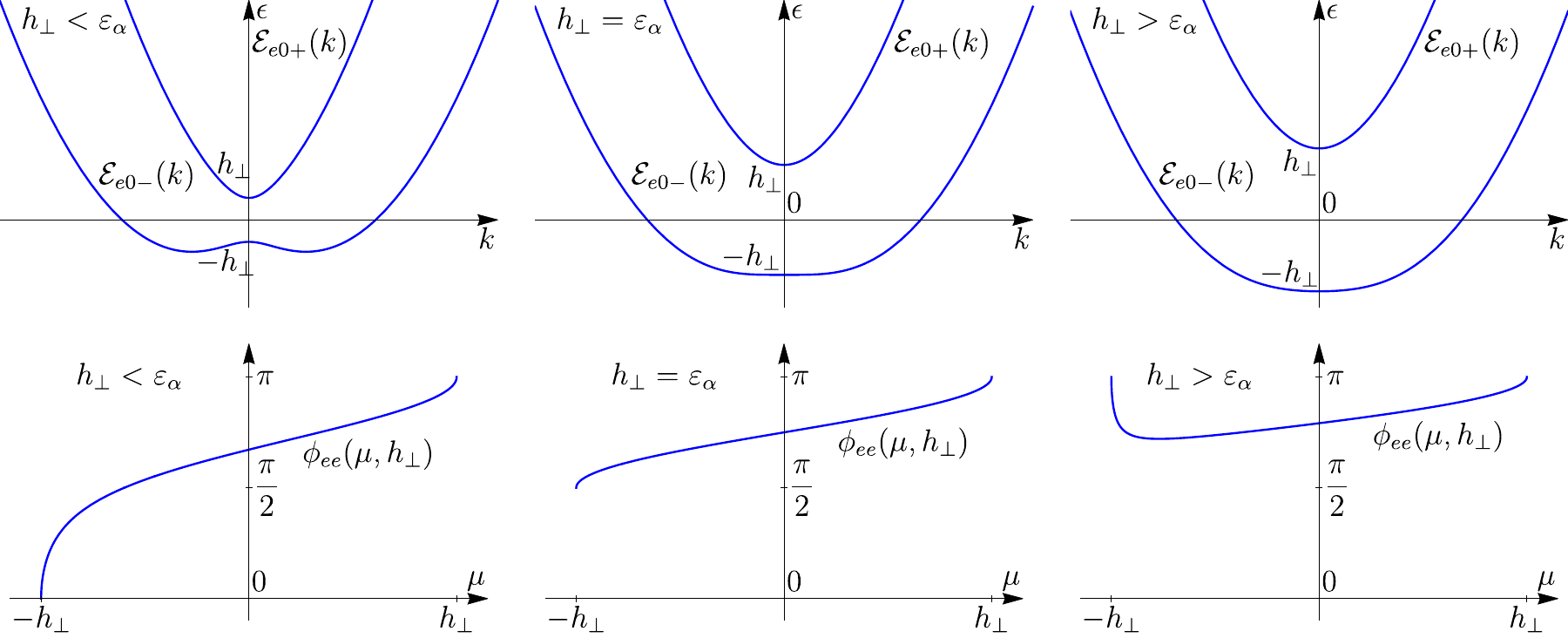}
\caption{
Two different regimes $h_\p \gtrless \eps_\al$ of the behavior
of the lower band $\Ec_{e0-}(k)$ [\eq{Ee0}] of the Hamiltonian $\Hc_{e0}(k)$ [\eq{qwHe0}] of the quantum wire at $h_z=0$
and the borderline case $h_\p=\eps_\al$ between them.
The upper row shows the spectra.
The lower row shows the dependence of the scattering phase $\phi_{ee}(\mu,h_\p)$ [\eq{qwUee}]
in the low-energy normal-reflection boundary conditions [\eqs{bcNR}{qwbce}]
on the chemical potential $\mu$ in each case.
}
\lbl{fig:qwphi}
\end{figure*}

The phase factor $U_{ee}(\mu,h_\p)$ is a dimensionless function
of the parameters of the microscopic model,
the chemical potential $\mu$ and the $h_\p$ Zeeman field
relative to the spin-orbit scale $\eps_\al$ [\eq{ea}].
In \figr{qwphi}, we plot the dependence of the scattering phase
$\phi_{ee}(\mu,h_\p)$, defined in \eq{qwUee},
on the chemical potential $\mu$.
The dependence differs qualitatively
in the two regimes $h_\p\gtrless\eps_\al$
of the behavior of the lower band $\Ec_{e0-}(k)$ [\eq{Ee0}].
At the upper bound $\mu=h_\p$ of the 1FS range $\mu\in(-h_\p,h_\p)$,
\[
    U_{ee}(\mu=h_\p,h_\p\gtrless\eps_\al)=-1
\]
in both regimes. At its lower bound $\mu=-h_\p$,
\[
    U_{ee}(\mu=-h_\p,h_\p\gtrless\eps_\al)=\mp1.
\]
Accordingly, for $h_\p<\eps_\al$, the phase $\phi_{ee}(\mu,h_\p)$ monotonically increases from $0$ to $\pi$
as $\mu$ spans $(-h_\p,h_\p)$.
For $h_\p>\eps_\al$, $\phi_{ee}(\mu,h_\p)$ has a minimum.
In the borderline case $h_\p=\eps_\al$,
\[
    U_{ee}(\mu=-h_\p,h_\p=\eps_\al)=-\ix,
\]
and $\phi_{ee}(\mu,h_\p=\eps_\al)$ monotonically increases from $\pi/2$ to $\pi$
as $\mu$ spans $(-h_\p,h_\p)$.

\section{Edge of a quantum spin Hall system \lbl{sec:qsh}}

In this section, we present another demonstration of the derivation of the low-energy model for 1FS with its BCs,
this time from the model for the edge of the quantum spin Hall system, considered in Ref.~\ocite{Fu}.
For simplicity, we assume that the Fermi level is close to the crossing point
of the counterpropagating edge states,
so that the normal-state electron Hamiltonian
for the two-component electron wave function
\[
    \Psih_e(x)=\lt(\ba{c} \Psi_{e+}(x) \\ \Psi_{e-}(x) \ea\rt)
\]
can already be taken linear in momentum, in the form
\beq
    \Hch_e(\kh)=\lt(\ba{cc} v\kh-\mu & 0 \\ 0  & -v\kh-\mu \ea\rt), \spc x > 0.
\lbl{eq:Heqsh}
\eeq

This Hamiltonian satisfies TR symmetry $\Tc_{e-}$ [\eq{Te}] with $\Th_{e-}=\ix\tau_y$
and, as discussed in \appr{T}, the edge states cannot be confined via backscattering without breaking it.
One way to confine the edge states of a quantum spin Hall system is,
as per the proposal of Ref.~\ocite{Fu}, by coupling it to a magnetic material.
The effect of the magnetic material, which we assume to be placed in the region $x<0$,
can be modeled by additional terms $m_{x,y}$ in the Hamiltonian \eqn{Heqsh}
\beq
    \Hch_e(\kh)
    =\lt(\ba{cc} v\kh-\mu & m_x-\ix m_y \\ m_x+\ix m_y & -v\kh-\mu \ea\rt),\spc x<0.
\lbl{eq:Heqshm}
\eeq
These terms open up a gap around the crossing point
for energies $\e+\mu\in(-m_\p,m_\p)$, $m_\p=\sq{m_x^2+m_y^2}$.

For the derivation of the low-energy model describing the energies $\e$ close to the Fermi level,
we consider the electron wave function
\[
    \Psih_e(x)=\lt(\ba{c} \psi_{e+}(x)\ex^{+\ix k_0x} \\ \psi_{e-}(x) \ex^{-\ix k_0x}\ea\rt), \spc x>0,
\]
as expansion in terms of the eigenstates at the Fermi level $\e=0$; $k_0=\mu/v$ is the Fermi momentum.
The coordinate-dependent expansion coefficients $\psi_{e\pm}(x)$
make up the low-energy wave function \eqn{psie},
which varies over the spatial scales much exceeding $v/\mu$ and $v/m_\p$ and is defined only in the region $x>0$.
The Hamiltonian for it in that region has the form $\Hh_e^{\Tc_{e-}}(\ph)$ [\eq{HeT}].

Due to the gap opening in the region $x<0$,
the wave function $\Psih_e(x)$ at energies $\e+\mu\in(-m,m)$ will decay as $x \rarr-\iy$
over a microscopic scale $v/\mu,v/m_\p$.
For the low-energy wave function $\psih_e(x)$, this will result in an effective BC at $x=0$.

Again, as in \secr{bcqw}, for the sake of deriving the BC,
it is sufficient to consider the energy $\e=0$ right at the Fermi level $\mu\in(-m_\p,m_\p)$.
We construct a general solution to the Schr\"ondinger equation $\Hch_e(\kh)\Psih_e(x)=\nm$
that decays into the magnetic region (as $x\rarr-\iy$) and does not grow into the bulk (as $x\rarr+\iy$).
In the region $x>0$, the general solution
\[
    \Psih_e(x)=\lt(\ba{c} \psi_{e+}\ex^{+\ix k_0x} \\ \psi_{e-} \ex^{-\ix k_0x}\ea\rt), \spc x>0,
\]
with constant coefficients $\psi_{e\pm}$, consists only of the particular solutions representing the low-energy wave function.

In the region $x<0$, the solution reads
\[
    \Psih_e(x)
    = C\lt(\ba{c} m \\ \ix\sq{m_\p^2-\mu^2}+\mu\ea\rt)\ex^{\f{\sq{m_\p^2-\mu^2}}{v}x}, \spc x<0,
\]
with $m=m_x-\ix m_y$ and an arbitrary coefficient $C$.

As follows from the differential properties of the Hamiltonian [\eqs{Heqsh}{Heqshm}],
the solution to the Schr\"odinger equation must be continuous at $x=0$, i.e., $\Psih_e(x=+0)=\Psih_e(x=-0)$.
This leads to two relations
\[
    \psi_{e+}= C m,\spc \psi_{e-}=C(\ix\sq{m_\p^2-\mu^2}+\mu)
\]
for the three coefficients $\psi_{e\pm}$ and $C$.
Excluding the coefficient $C$ from these equations, we obtain a relation
\beq
    \psi_{e+}=U_{ee}\psi_{e-}
\lbl{eq:qshbce}
\eeq
with
\[    U_{ee}
    =\f{m}{\ix\sq{m_\p^2-\mu^2}+\mu}.
\]
Again, the relation \eqn{qshbce} represents the BC for the low-energy electron wave function
and $U_{ee}$ is the phase factor therein.
The BC for the hole part $\psih_h(x)$ of the low-energy BdG wave function
can be obtained analogously.
We thus  find that indeed the BCs for the low-energy model derived
from the model of the edge of the quantum spin Hall system coupled to a magnetic material
have the form \eqn{bcNR} of the general normal-reflection BCs.

\section{Low-energy symmetry-based approach and topology, outlook
\lbl{sec:conclude}}

The main physical finding of this work
and its practical implications have already been formulated and discussed in \secsr{intro}{bs}.
In this concluding section, we briefly discuss
the relation of the employed theoretical formalism
to the topological aspect of the system.

The low-energy symmetry-based formalism
for studying the bound states in topological systems
consists of two stages: (i)
deriving the low-energy model of the most general form,
whose Hamiltonian and BCs are subject only to symmetries and the fundamental principle of probability-current conservation;
(ii) calculating and exploring the corresponding bound-state structure of the model.
As such, remarkably, this formalism allows one
to obtain and explore generic bound-states structures of topological systems
without ever explicitly invoking the notion of topology.
Nonetheless, the so-obtained bound-state structures
will, of course, be in accord with the topological properties of the systems.
For the particular class of systems studied in this paper,
quasi-1D superconductors with odd number of FSs, interfaced with a vacuum or an insulator,
the obtained ever-present Majorana bound state indeed most likely has topological origin.
The $\e=0$ bound states {\em are} topologically protected under CC symmetry $\Cc_+$, i.e.,
cannot be removed without breaking the symmetry or closing the bulk gap.
In topological systems, bound states are anticipated based on the concept of bulk-boundary correspondence~\cite{Chiu},
which, however, to the best of our knowledge, remains unproven at the required level of rigor.
Regardless, even if such proof is possible,
the low-energy approach has apparent advantages
in that it delivers the generic bound-state structure, which can be explored in an explicit fashion.
As such, the formalism can be used not only to confirm or illustrate topological concepts,
but rather, to test them and possibly discover new features.

This low-energy symmetry-based formalism
is completely general and applicable to a multitude of other systems
(as long as they allow for a well-defined low-energy limit).
It has previously been applied to 2D chiral-symmetric semimetals~\cite{KharitonovLSM}
and quantum anomalous Hall systems (Chern insulators)
in the vicinity of the topological phase transition~\cite{KharitonovQAH}.
Regarding superconductors, possible further extensions of the present work could be
exploring the meaning and possible physical realizations of Andreev-reflection BCs for 1FS,
extending the approach to higher dimensions.
The low-energy model for two Fermi surfaces should, in particular, be useful
for studying bound states in superconductors with magnetic interfaces or magnetic scatterers~\cite{Rouco1,Rouco2}.

\section{Relation to previous work \lbl{sec:Samokhin}}

During the preparation of the manuscript, Ref.~\ocite{Samokhin}
came out, where a similar main conclusion about the existence of Majorana bound states
was reached using a similar low-energy model.
The main differences between our work and Ref.~\ocite{Samokhin} are as follows.
We derive the low-energy model based purely on CC symmetry
and the current conservation principle, thereby proving that it is of the most general form.
Whereas in Ref.~\ocite{Samokhin}, neither CC symmetry nor the current conservation principle were addressed.
The normal-reflection BCs were not derived, but rather postulated phenomenologically,
and the Andreev-reflection BCs for the case of 1FS did not arise in Ref.~\ocite{Samokhin} at all.
Further, the effective TR symmetry $\Tc_{e+}$ of the electron Hamiltonian
was assumed necessary in Ref.~\ocite{Samokhin}, which prohibited the Fermi-point mismatch.
We take the Fermi-point mismatch into account
(as well as the coordinate dependence of the pairing field that helps mitigate it)
and demonstrate that Majorana bound states persist in its presence,
which is an important finding for practical applications.

\begin{acknowledgments}
%\section*{Acknowledgements}

M.K. acknowledges financial support by the DFG Grant No. KH 461/1-1.
E.M.H. and B.T. acknowledge funding by the Deutsche Forschungsgemeinschaft
(DFG, German Research Foundation) through SFB 1170,
project-id 258499086, through Grant No. HA 5893/4-1 within SPP 1666
and through the W\"urzburg-Dresden Cluster of Excellence on Complexity
and Topology in Quantum Matter – ct.qmat (EXC 2147, project-id 390858490),
as well as by the ENB Graduate School on Topological Insulators.
The work of F.S.B. was partially funded by Spanish Ministerio de Ciencia,
Innovacion y Universidades (MICINN) [Projects FIS2017-82804-P and PID2020-114252GB-I00 (SPIRIT)],
and by Grupos Consolidados UPV/EHU del Gobierno Vasco (Grant No.  IT1249-19).

\end{acknowledgments}

\appendix
\section{Time-reversal symmetries \lbl{app:T}}

In this Appendix,  we consider the effects of possible additional TR
symmetries $\Tc_\pm$ of the low-energy model for 1FS.

\subsection{Time-reversal symmetries of the Hamiltonian \lbl{app:TH}}

When introduced formally, a TR symmetry of a Hamiltonian means
that there exists an anti-unitary TR operation
\beq
    \Tc_\pm=\Th_\pm\Kc,\spc
    \Tc_\pm\psih(x)=\Th_\pm\psih^*(x),
\lbl{eq:Tpsi}
\eeq
under which the Hamiltonian remains invariant,
\beq
    \Th_\pm [\Hh(\ph)]^* \Th_\pm^\dg=\Hh(\ph),
\lbl{eq:TH}
\eeq
with $\ph^*=-\ph$.
Here, $\Th_\pm$ are unitary matrices.
Two different types $\Tc_\pm$ of TR operation are possible,
squaring to either plus or minus unity,
\[
    \Tc^2_\pm
    =\Th_\pm\Th_\pm^*=\pm\um,
\]
which we label accordingly with $\pm$.

For a spinful electron system, the actual TR operation is of the type $\Tc_-$.
However, an effective $\Tc_+$ symmetry could also be present in a spinful electron system,
which can arise as a combination of the actual $\Tc_-$ and some spatial operation,
as is the case, e.g., for the quantum-wire system, see \figr{qw} and \appr{qwT}.

We construct the most general forms of the $\Tc_\pm$ operations for the low-energy Hamiltonian $\Hh(\ph)$ [\eq{H}].
The TR operation, whether $\Tc_+$ or $\Tc_-$,
interchanges the right- and left-moving electron states $\psi_{e\pm}(x)$.
The most general forms of the TR operations $\Tc_{e\pm}=\Th_{e\pm}\Kc$
acting only on the electron part $\psih_e(x)$  of the BdG wave function \eqn{psi}  are
\beq
    \Th_{e+}=\tau_x,\spc
    \Th_{e-}=\ix\tau_y,
\lbl{eq:Te}
\eeq
\beq
    \Tc_{e\pm}\psih_e(x)
    =\lt(\ba{c}\psi_{e-}^*(x)\\\pm\psi_{e+}^*(x)\ea\rt).
\lbl{eq:Tepsi}
\eeq
The overall phase factors of the operations can be chosen arbitrarily.
For the $\Tc_{e-}$ operation, the electron states $\psi_{e\pm}(x)$ form a Kramers pair.

Either of these TR symmetries $\Tc_{e\pm}$
(to avoid confusion, here and below, we mean one of the operations at a time, but join them into one formula)
restricts the form of the electron Hamiltonian $\Hh_e(\ph)$ [\eq{He}] to
\beq
    \Hh_e^{\Tc_{e\pm}}(\ph)=\lt(\ba{cc} v\ph & 0 \\ 0 & -v\ph \ea\rt),
\lbl{eq:HeT}
\eeq
i.e., makes the Fermi velocities equal, $v_+=v_-=v$,
and prohibits the Fermi-point mismatch,
$k_+=k_-=k_0$, $\eps_0=0$.

When combined with the CC operation $\Cc_+$ [\eq{C+}],
the TR operation $\Tc_{e\pm}$ [\eq{Te}] for the electron part $\psih_e(x)$ of the wave function
naturally introduces the TR operation $\Tc_{h\pm}=\Th_{e\pm}^*\Kc$ for the hole part $\psih_h(x)$.
These electron $\Tc_{e\pm}$ and hole $\Tc_{h\pm}$ parts
can be joined into the total TR operation
\beqar
    \Tc_\pm
    &=&\lt(\ba{cc} \Th_{e\pm} & \nm \\ \nm & a_\pm \Th_{e\pm}^* \ea\rt)\Kc,\nn\\
    \Th_+&=&\lt(\ba{cc} \tau_x & \nm \\ \nm & a_+\tau_x\ea\rt),
    \spc
    \Th_-=\lt(\ba{cc} \ix\tau_y & \nm \\\nm & a_-\ix\tau_y\ea\rt),
\lbl{eq:T}
\eeqar
acting on the full BdG wave function $\psih(x)$ [\eq{psi}],
where, importantly, $a_\pm$ is an arbitrary relative phase factor, $|a_\pm|=1$.

Applying this TR operation to the Hamiltonian \eqn{H},
we obtain that the most general form of the Hamiltonian
satisfying the TR symmetry $\Tc_\pm$ [\eqs{TH}{T}]
has the form
\beq
    \Hh^{\Tc_\pm}(\ph)=
    \lt(\ba{cccc}
        v\ph & 0 &0&\De(x)\\
        0&-v\ph &-\De(x)&0\\
        0&-\De^*(x)& v\ph&0\\
        \De^*(x)&0&0&-v\ph
    \ea\rt),
\lbl{eq:HT}
\eeq
where the pairing field $\De(x)$ and the relative phase factors $a_\pm$ must satisfy the constraint
\beq
    \De(x)=\mp\De^*(x)a^*_\pm.
\lbl{eq:Deaconstr}
\eeq
Hence, the pairing field must be a real function $\De'(x)$, up to a constant phase factor,
\beq
    \De(x)=\De'(x)\ex^{-\ix\de}
\lbl{eq:DeT}
\eeq
and the phase factors in \eq{T} must satisfy
\beq
    a_\pm=\mp\ex^{2\ix\de}.
\lbl{eq:a}
\eeq

In other words, to satisfy either of the symmetries $\Tc_\pm$,
the pairing field must be effectively real: the overall constant phase factor $\ex^{-\ix\de}$
is physically inessential, since it can be adjusted by the phase factors of the basis functions of $\psih_e(x)$ and $\psih_h(x)$ parts,
and so, it cannot affect the symmetry.
The real function $\De'(x)$ could, in principle, change sign, which would create a domain-wall structure.

In particular, under $\Tc_\pm$ symmetry, the one-harmonic coordinate dependence \eqn{Deq}
of the pairing field is prohibited for finite momentum $q\neq 0$,
which is in accord with the fact that under $\Tc_\pm$ the Fermi-point mismatch is also prohibited.
This dependence was introduced in \secr{Deq}
as a possible mechanism to mitigate the effect of the Fermi-point mismatch.
When the latter is absent, there is no practical reason for this form of the pairing field.
On the other hand, the constant pairing field $\De(x)=\De_0$
in the low-energy 1FS model is always $\Tc_\pm$-symmetric.

We note an interesting property that if the low-energy Hamiltonian $\Hh(\ph)$ satisfies {\em one of} the $\Tc_\pm$ symmetries,
i.e., \eqs{HT}{DeT} hold, then it also automatically satisfies the other symmetry $\Tc_\mp$, respectively.
In this sense, the other symmetry $\Tc_\mp$ has {\em emergent} character.
At the same time, we also note that this symmetry relation is limited
since, even though it holds for the bulk Hamiltonian, the properties of the BCs
under $\Tc_\pm$ symmetries are radically different, as we show next.

\subsection{Time-reversal symmetries of the normal-reflection boundary conditions \lbl{app:Tbc}}

Now we turn to the TR symmetries $\Tc_{\pm}$ of the normal-reflection BCs \eqn{bcNR}.
As explained in \secr{bc}, the BCs satisfy a certain symmetry if the wave function
transformed by the symmetry operation also satisfies those BCs.
As with $\Cc_+$, TR operations $\Tc_\pm$ do not alter the coordinate and thus leave the geometry of the system intact;
hence, the system with a boundary can, in principle, be $\Tc_\pm$-symmetric.
Inserting $\Tc_\pm\psih(0)$ [\eqs{Tpsi}{T}]
into the normal-reflection BCs \eqn{bcNR} with $v_+=v_-$, as required by $\Tc_\pm$ symmetry of the Hamiltonian \eqn{HT},
we find that the BCs are $\Tc_\pm$-symmetric if
\[
    U_{ee}=\pm U_{ee},
\]
respectively.

Hence, the normal-reflection BCs \eqn{bcNR} with $v_+=v_-$ and arbitrary phase factor $U_{ee}$, $|U_{ee}=1|$, satisfy $\Tc_+$ symmetry.

On the other hand, interestingly, we obtain that there are {\em no normal-reflection BCs} that satisfy $\Tc_-$ symmetry.
This result is the manifestation of the known physical effect:
absence of backscattering between the counterpropagating 1D electron states of one Kramers pair $\psi_{e\pm}(x)$
(this effect is not related to superconductivity).
One physical realization of such system with the low-energy Hamiltonian \eqn{HeT}
is the edge of a quantum spin Hall system in the topologically nontrivial phase~\cite{BHZ}.
The counterpropagating edge states go around the whole closed boundary of a 2D finite-size sample.
Cutting the sample into parts will not cause backscattering; rather, the edge states will continue
to propagate along the newly created edges of the parts.
Thus, nonexistence of BCs for a $\Tc_-$-symmetric system of one Kramers pair with the Hamiltonian \eqn{HeT}
has a natural physical explanation and it is quite remarkable that
the presented formalism of general BCs is ``aware'' of such physical effects.

In order to create an inaccessible region $x<0$
in such $\Tc_-$-symmetric quasi-1D system
from which the states can backscatter, $\Tc_-$ symmetry must necessarily be broken.
In practice, this can be achieved by bringing a magnetic material in contact with the quantum-spin-Hall system
(\figr{qsh}), which induces a gap in part of its edge, the situation we consider in \secr{qsh}.

The BdG systems with CC symmetry $\Cc_+$ and additional TR symmetries $\Tc_\pm$
belong to BDI and DIII symmetry classes, respectively.
Note that the presence of $\Cc_+$ and $\Tc_\pm$ symmetries automatically
also generates chiral symmetries with the operations $\Tc_\pm \Cc_+$.
These, however, do not have a profound effect for the case of 1FS model.

As far as the consequences of the $\Tc_\pm$ symmetries for the Majorana bound states,
we see that the additional $\Tc_+$ symmetry does have an effect on the bulk, prohibiting the Fermi-point mismatch,
but does not directly affect the Majorana bound state,
which exists in the gapped superconducting state, regardless of whether $\Tc_+$ is present or not.
On the other hand, it is simply impossible to create a boundary without breaking the
TR symmetry $\Tc_-$.

\subsection{Time-reversal symmetries of the generalized quantum-wire model \lbl{app:qwT}}

We now consider the TR symmetries $\Tc_\pm$ of the generalized quantum-wire model of \secr{qw}
and establish the relation between them and those of the low-energy model, to which it reduces.

The actual TR symmetry $\Tc_{e-}=\ix\sig_y\Kc$ of electrons
with the Hamiltonian $\Hch_e(\kh)$ [\eq{qwHe}]
is broken by the Zeeman field, which transforms as
\[
    \Tc_{e-}:\spc (h_x,h_y,h_z)\rarr-(h_x,h_y,h_z).
\]
However, as was noticed in Refs.~\ocite{Sprb95,Sann,Samokhin},
the electron Hamiltonian $\Hch_{e0}(\kh)$ [\eq{qwHe0}] with only the $h_\p$ Zeeman field
possesses an effective TR symmetry $\Tc_{e+}=\sig_x\Kc$.
This TR operation arises as the product $\Tc_{e+}=\Sig_z\Tc_{e-}$ of the actual TR operation $\Tc_{e-}$
and the reflection $\Sig_z$ along the horizontal direction perpendicular to the wire, \figr{qw}.
The Zeeman field transforms as
\[
    \Sig_z:\spc (h_x,h_y,h_z)\rarr (-h_x-h_y,h_z)
\]
under the latter and, hence, transforms as
\[
    \Tc_{e+}:\spc (h_x,h_y,h_z)\rarr (h_x,h_y,-h_z)
\]
under the effective $\Tc_{e+}$.
Thus, the components $h_{x,y}$ of the Zeeman field in the vertical $xy$ plane containing the wire
are preserved under $\Tc_{e+}$, even though they are not preserved under either of these two operations individually.
Importantly, the electron system {\em with a boundary} satisfies $\Tc_{e+}$;
and indeed, the hard-wall BCs \eqn{hwbce} satisfy $\Tc_{e+}$.

The effective TR symmetry $\Tc_{e+}$ of the electron part of the microscopic quantum-wire model [\eqs{qwHe0}{hwbce}]
with $h_z=0$ translates to that [\eq{Te}] of the low-energy model,
in accord with \appr{TH}:
it prohibits the Fermi-point mismatch [$\eps_0=0$, $k_+=k_-=k_0$, \eq{qwe0}],
enforces equal velocities [$v_+=v_-=v$, \eq{qwv}],
but provides no constraints on the scattering phase factor $U_{ee}$ [\eq{qwUee}] of the normal-reflection BCs \eqn{bcNR}.

Further, as for the low-energy model (\appr{TH}),
the general forms of the TR operations for the BdG Hamiltonian \eqn{qwH} of the wire read
\beqar
    \Tc_\pm
    &=&\lt(\ba{cc} \Th_{e\pm} & \nm \\ \nm & \at_\pm \Th_{e\pm}^* \ea\rt)\Kc,\nn\\
    \Th_+&=&\lt(\ba{cc} \tau_x & \nm \\ \nm & \at_+\tau_x\ea\rt),
    \spc
    \Th_-=\lt(\ba{cc} \ix\tau_y & \nm \\\nm & \at_-\ix\tau_y\ea\rt),
\lbl{eq:qwT}
\eeqar
with adjustable phase factors $\at_\pm$, $|\at_\pm|=1$.
Applying these operations, we find that the pairing field \eqn{qwDe} is $\Tc_+$-symmetric
if its components and $\at_+$ satisfy
\[
    \De_{0,z}(x;k)
    %=-\at_+\De_0^*(X,-k)
    =-\at_+^*\De_{0,z}^*(X,k),\spc
    %\nn\\
    \De_{x,y}(x;k)
    %=-\at_+\De_{x,y}^*(X,-k)
    =\at_+^*\De_{x,y}^*(X,k),\spc
    %\nn\\
%    \De_z(x;k)
    %=+\at_+\De_z^*(X,-k)
%    =-\at_+^*\De_z^*(X,k).
\]
(note that different triplet components transform differently;
this is in accord with $\Tc_{e+}=\Tc_{e-}\Sig_z$ involving a spatial operation,
so the spin orientation of the pairing field does matter).
These relations are satisfied when
\beqar
    \De_{0,z}(x;k)&=&\ex^{-\ix\de}\De_{0,z}'(x;k),\nn\\
    \De_{x,y}(x;k)&=&\ex^{-\ix\de}\ix\De_{x,y}'(x;k),\nn\\
    \ex^{2\ix\de}&=&-\at_+,\nn
\eeqar
where $\De_{0,x,y,z}'(x;k)$ are real.
Inserting this form into \eq{qwDele} for the low-energy pairing field $\De(x)$,
we find that the latter does have the form \eqn{DeT} required for it to be $\Tc_+$-symmetric.

Similarly, the pairing field is $\Tc_-$-symmetric when
\[
    \De_{0,x,y,z}(x;k)
    %=-\at_+\De_{x,y}^*(X,-k)
    =\at_-^*\De_{0,x,y,z}^*(x;k).
\]
This holds when
\[
    \De_{0,x,y,z}(x;k)=\ex^{-\ix\de}\De_{0,x,y,z}'(x;k)
,\spc    \ex^{2\ix\de}=\at_-,
\]
where $\De_{0,x,y,z}'(x;k)$ are real.
Inserting this form into \eq{qwDele} for $\De(x)$,
we find that the latter does not generally have the form \eqn{DeT} required for it to be $\Tc_-$-symmetric.
This is not surprising since, even if the microscopic pairing field is $\Tc_-$-symmetric,
the electron Hamiltonian $\Hch_{e0}(\kh)$ [\eq{qwHe0}] is not, and it is involved in the low-energy projection \eqn{qwDele}.

These results and those of \appr{TH} illustrate the general property
that the symmetries of the microscopic model
and of the low-energy model to which the former reduces are not necessarily identical.
If the microscopic model possesses some symmetries, then surely the low-energy
model does also, which is the case for $\Tc_+$ symmetry here.
However, the low-energy model can possess some additional, {\em emergent} symmetries
that are not present in the microscopic model from which it originates.
This is the case for $\Tc_-$ symmetry here (and this applies only to the low-energy bulk Hamiltonian, but not to the BCs).
At the end of \appr{TH},  we noted that if the low-energy BdG Hamiltonian satisfies $\Tc_+$ symmetry [\eq{HT}]
then it also satisfies $\Tc_-$ symmetry (and vice versa). However, in the microscopic model, $\Tc_-$ symmetry is broken.


\begin{thebibliography}{99}
\bibitem{Te}
%a footnote about $\Tc_{e\pm}$ label
$\Tc_{e\pm}$ with the label $e$
will denote the symmetries of just the electron system,
whereas $\Tc_{\pm}$ will denote the symmetries of the whole BdG system if electrons and holes,
in the presence of superconductivity.
This distinction is useful, since
the superconducting pairing field could, in principle, break these symmetries,
see discussion in \appr{T}.
\bibitem{micro}
We will refer to these models  as ``microscopic'' in relation to the derived low-energy model
in the sense that they contains more degrees of freedom
or the structure of their boundary is specified.
These models, however, could themselves arise as low-energy models from other, ``more microscopic'' models.
\bibitem{k0}
Note that the Fermi points $\pm k_0$ [\eq{qwk0}] of the
Hamiltonian $\Hch_{e0}(\kh)$ without $h_z$ are actually those symmetric crossing points \eqn{e0}
of the electron and hole normal-state bands
in the presence of (small) $h_z$, compare \figsr{qw}{system}.
\bibitem{hwbce}
We mention that for the continuum quantum-wire model one could, in principle,
also derive and study the most general form of the BCs;
here, we consider the familar hard-wall BCs as an example,
which are just one instance of the latter.
\end{thebibliography}
\end{document}